\documentclass[10pt]{article}

\usepackage[margin=1in]{geometry}
\usepackage{amsmath,amssymb}
\usepackage{booktabs}
\usepackage{graphicx}
\usepackage{subcaption}
\usepackage{xcolor}
\usepackage{microtype}
\usepackage{siunitx}
\usepackage{float}
\usepackage{authblk}

\usepackage[
    colorlinks=true,
    linkcolor=black,
    citecolor=blue,
    urlcolor=blue
]{hyperref}

\newcommand{\pt}{p_{\mathrm{T}}}
\newcommand{\TrackML}{\textsc{TrackML}}

\title{%
AlphaHits: Full-Event Tracking and Reconstruction\\ Towards a Hit-Level Foundation Model}

\author[1,2]{Sascha Caron}
\author[1,2,3]{Nadezhda Dobreva}
\author[1,2,$\dagger$]{Polina Moskvitina}
\author[1,2,3]{Eugene Shalugin}
\author[4]{Javad Sohrabi}
\author[4]{Roberto Ruiz}
\author[5]{Yue Zhao}
\affil[1]{National Institute for Subatomic Physics (Nikhef), Amsterdam, The Netherlands}
\affil[2]{High-Energy Physics, Radboud University, Nijmegen, The Netherlands}
\affil[3]{CERN, Geneva, Switzerland}
\affil[4]{Instituto de F\'{\i}sica Corpuscular, CSIC-Universitat de Val\`encia, E-46980 Paterna, Valencia, Spain}
\affil[5]{SURF, Amsterdam, The Netherlands}
\affil[$\dagger$]{Current affiliation: Argonne National Laboratory, USA}

\date{}

\begin{document}
\maketitle

\begin{abstract}
We introduce AlphaHits, a Transformer-based approach towards hit-level foundation models for LHC physics, building on the TrackFormer and HiggsFormer architectures. AlphaHits operates directly on detector hits and demonstrates scalable full-event charged-particle reconstruction together with multi-task event reconstruction and classification from low-level detector information.

For charged-particle tracking, AlphaHits-TrackFormer reconstructs the full charged-particle content of complete TrackML events without imposing a transverse-momentum selection on the target tracks. The model processes both pixel and strip detector hits in events containing up to \(1.5\times10^5\) hits under HL-LHC-like pileup conditions and assigns the detector hits to individual reconstructed trajectories across the full kinematic range represented in the dataset. This includes the highly challenging low-\(p_T\) population, extending down to approximately \(150\,\mathrm{MeV}\) in the supplied TrackML samples. 
AlphaHits-TrackFormer achieves a TrackML score of 0.959 when evaluated inclusively over all reconstructable tracks, representing, to our knowledge, the highest reported TrackML score for full-detector, inclusive charged-particle reconstruction.

Building on the TrackFormer architecture, AlphaHits is
pretrained on charged-particle tracking and fine-tuned for
multi-task reconstruction and classification. From a shared
hit-level representation, it predicts tracks, jets, jet
flavour, and event-level observables. AlphaHits achieves an AUC of 0.944 for Higgs event classification, demonstrating performance comparable to that of a Particle Transformer (ParT) baseline using fully reconstructed inputs. It also achieves
performance comparable to Delphes-based reconstruction
baselines for jet reconstruction, jet flavour classification,
and missing transverse momentum regression.

Together, these results highlight the potential for scalable full-event tracking and multi-task learning directly from detector hits, representing a promising step towards transferable hit-level representations and general-purpose foundation models for LHC physics.

\end{abstract}

\section{Introduction}
Particle reconstruction at the Large Hadron Collider traditionally relies on a sequence of specialized algorithms that transform low-level detector measurements into tracks, vertices, jets, flavour-tagging information, and other reconstructed physics objects before event-level inference is performed. Modern machine-learning approaches have begun to replace individual components of this chain, but tracking studies commonly focus on a restricted subset of charged particles, often imposing transverse-momentum thresholds around \(p_T\sim1~\mathrm{GeV}\). Such selections simplify the reconstruction problem, but exclude a large fraction of the charged-particle population: a cut at \(p_T\sim1~\mathrm{GeV}\) cuts more than 90 percent of the particles.

The question behind this work is whether a learned representation can instead be constructed from the \emph{full detector event}, retaining information from the complete charged-particle population rather than restricting reconstruction to a high-\(p_T\) subset. AlphaHits-TrackFormer therefore processes the full pixel and strip hit cloud and attempts to reconstruct all reconstructable charged particles represented in the TrackML sample, without imposing a \(p_T\) threshold on the target tracks. This substantially increases the combinatorial difficulty of the tracking problem, particularly in the dense low-\(p_T\) regime, but also preserves information that would otherwise be discarded before higher-level event analysis.

We further ask whether such hit-level representations can support tasks beyond charged-particle tracking. AlphaHits extends this approach to multi-task event reconstruction and classification, exploring whether a common representation of low-level detector information can be reused across reconstruction and inference tasks. The combination of inclusive detector-level representation learning and transfer across physics objectives constitutes an important step towards a hit-level foundation model for collider physics. This direction aligns with the broader vision of Large Physics Models, which aims to develop general-purpose representations transferable across diverse tasks in high-energy physics~\cite{barman2025largephysicsmodels}.

Our earlier TrackFormer and HiggsFormer studies explored this idea from complementary directions: TrackFormer investigated Transformer-based assignment of inner-tracker hits to charged-particle trajectories~\cite{Caron2025Trackformers}, and ``TrackFormers Part~2" introduced a sparse attention so that larger hit collections fit in memory  ~\cite{Caron2025Trackformers2}. HiggsFormer showed that a related architecture could be retrained to classify Higgs events directly from raw inner-tracker hits, bypassing explicit track and object reconstruction~\cite{Caron2025Higgsformer}. Trained on the raw inner-tracker hits of simulated $t\bar{t}H (H\rightarrow b\bar{b})$ and $t\bar{t}$ events, it reached AUC of 0.855, comparable to what a Particle Transformer~\cite{Qu2022ParticleTransformer} on Delphes tracker objects reaches at the 40\% b-tagging working point (0.852).

This result suggested that models trained on low-level inputs can learn detector-hit representations which retain substantial information relevant not only to particle reconstruction but also to downstream physics inference.

In this work, we substantially extend both approaches within the AlphaHits framework, combining scalable full-event tracking with multi-task learning from shared hit-level representations.
\begin{itemize}
    \item AlphaHits-TrackFormer (our version 3 of the  TrackFormer idea) reconstructs tracks in complete TrackML events~\cite{Kiehn2019TrackMLKaggle} of up to $1.5\times10^5$ hits. A learned Router sparsifies the attention, a global Tracker model predicts binned and continuous track parameters for every hit, a charge-aware preconditioner and a learned Gate select the ambiguous regions, and a local Refiner resolves them. Trained on $7\,830$ events, it reaches TrackML scores of $0.959$ over all tracks ($p_T>0.15$ GeV).
    \item AlphaHits extends the earlier HiggsFormer approach, building on the AlphaHits-TrackFormer architecture. Its backbone is first trained on tracking alone. Nine heads are then trained jointly on the same hits and reconstruct tracks, jets with flavour tags, $H_T$, missing transverse momentum, leptons and the event class in one forward pass. On the $t\bar{t}H (H\rightarrow b\bar{b})$ vs $t\bar{t}$ hard scatter task AlphaHits reaches an AUC of 0.932 from silicon hits alone and 0.944 with calorimeter towers as an additional input. On the same events, a Particle Transformer on Delphes objects reaches 0.877 with the tracker alone and 0.924 with tracker and calorimeter.
\end{itemize}

Taken together, these results point towards a common paradigm in which scalable sparse neural networks learn physically informative representations directly from detector measurements, while task-specific objectives, heads, or local refinement stages extract reconstruction- and analysis-level information. Track reconstruction, event reconstruction and Higgs classification provide complementary demonstrations of this approach and motivate the development of more general hit-level Foundation Models for collider physics.

\section{Related Work}
Foundation models in high-energy physics have predominantly relied on reconstructed, high-level physics objects, such as jets and leptons, or particle-level jet constituents, demonstrating promising transferability across downstream analysis tasks \cite{golling2024mpm,vigl2024finetuning,hsu2026evenet,birk2024omnijet,mikuni2024omnilearn,bhimji2025omnilearned,amram2024aspen}. Recent advances have extended this paradigm towards lower-level detector representations, including calorimeter cells, clusters, and reconstructed tracks, highlighting the potential of unified foundation models for detector simulation and particle reconstruction \cite{kakati2026prompting}. Recent work on machine-learned particle-flow reconstruction
has further demonstrated the potential of learning reusable
representations from detector-level information, with
applications to jet tagging, energy regression, and missing
transverse momentum reconstruction
\cite{mokhtar2026mlpf}.

Recent work has explored unified reconstruction directly
from detector measurements in simulated electron--positron
collisions at FCC-ee \cite{hart2026rem}, with typical
event sizes of a few thousand detector hits.
In contrast, the HL-LHC presents substantially higher
detector occupancies and pileup, requiring models capable
of processing complete events with more than $10^5$ hits.
Scaling transferable hit-level representations to such
conditions remains an important challenge.

Machine-learning approaches to charged-particle tracking have increasingly focused on graph-based and end-to-end formulations. The TrackML challenge provided a common benchmark for large-scale tracking algorithms~\cite{Kiehn2019TrackMLKaggle}, while subsequent work demonstrated strong performance with graph neural networks and related geometric deep-learning approaches~\cite{Ju2021ExaTrkX,Biscarat2021RealisticGNNTracking,Caillou2024GNN4ITk}. Object-condensation methods provide an alternative formulation in which reconstruction is treated as learned object formation rather than explicit edge classification~\cite{Kieseler2020ObjectCondensation,Lieret2023HighPileupOC}. EggNet uses graph attention with an evolving graph structure to
reconstruct tracks directly from detector
hits~\cite{Calafiura2024EggNet}, with a subsequent study evaluating
its physics and computing performance on the full TrackML
dataset~\cite{Chan2025EggNetPerformance}. 
More recently, Transformer-based approaches have also been explored for charged-particle tracking, including TrackFormer-based models~\cite{Caron2025Trackformers,Caron2025Trackformers2}, TrackSorter~\cite{Melkani2024TrackSorter}, and other Transformer formulations~\cite{VanStroud2025TransformerTracking}. In contrast to approaches based on explicit hit graphs or sequential track construction, the method presented here combines sparse full-event parameter inference with a second learned local refinement stage to resolve residual ambiguities in track assignment.

HEPTv2~\cite{miao2026heptv2} reports a double-majority
tracking efficiency of 98.6\% with a 0.8\% fake rate,
but considers a substantially restricted reconstruction
problem. It uses only pixel-detector hits and targets
particles with $p_T>0.9$~GeV, excluding approximately
90\% of charged particles from the reconstruction targets.
Hits from excluded particles remain in the input but
are treated as background rather than reconstructed
into separate tracks. In contrast, AlphaHits-TrackFormer
processes the complete pixel and strip detector,
reconstructing tracks down to $p_T>0.15$~GeV in events
containing up to $1.5\times10^5$ hits.

Scaling attention to detector events containing $\mathcal{O}(10^5)$ hits requires avoiding dense all-to-all self-attention. 
Memory-efficient attention implementations such as FlashAttention-2 ~\cite{Dao2023FlashAttention2ICLR} reduce the overhead of attention computation, while FlexAttention ~\cite{He2024FlexAttention} and structured sparse-attention methods make it possible to restrict communication to physically or algorithmically selected subsets of the event. Related work has also explored efficient point Transformers based on locality-sensitive hashing for high-energy-physics applications~\cite{Miao2024HEPT}. Our approach builds on the sparse-attention strategy introduced in TrackFormers Part~2~\cite{Caron2025Trackformers2}, while extending it to learned routing and to both reconstruction and event-classification tasks. The local track refiner is additionally related in spirit to permutation-invariant object representations and slot-based prediction~\cite{Locatello2020SlotAttention,Cheng2022Mask2Former}, although here the slots represent local charged-particle hypotheses rather than image objects.

At the event level, Transformer architectures have achieved strong performance for jet and particle-based classification~\cite{Qu2022ParticleTransformer,Builtjes2025PhysicsInteractions}. HiggsFormer extended this direction to classification directly from raw inner-tracker hits~\cite{Caron2025Higgsformer}, demonstrating that substantial event-level physics information can be extracted without first reconstructing the conventional hierarchy of tracks, jets, and flavour-tagged objects.
The present work further scales this detector-level approach and augments it with multiple auxiliary reconstruction objectives, allowing the same broad architecture to be studied across both low-level tracking and high-level Higgs classification.

\section{Detector-Level Transformer Architectures}

\begin{figure}[H]
    \centering
    \includegraphics[width=\linewidth]{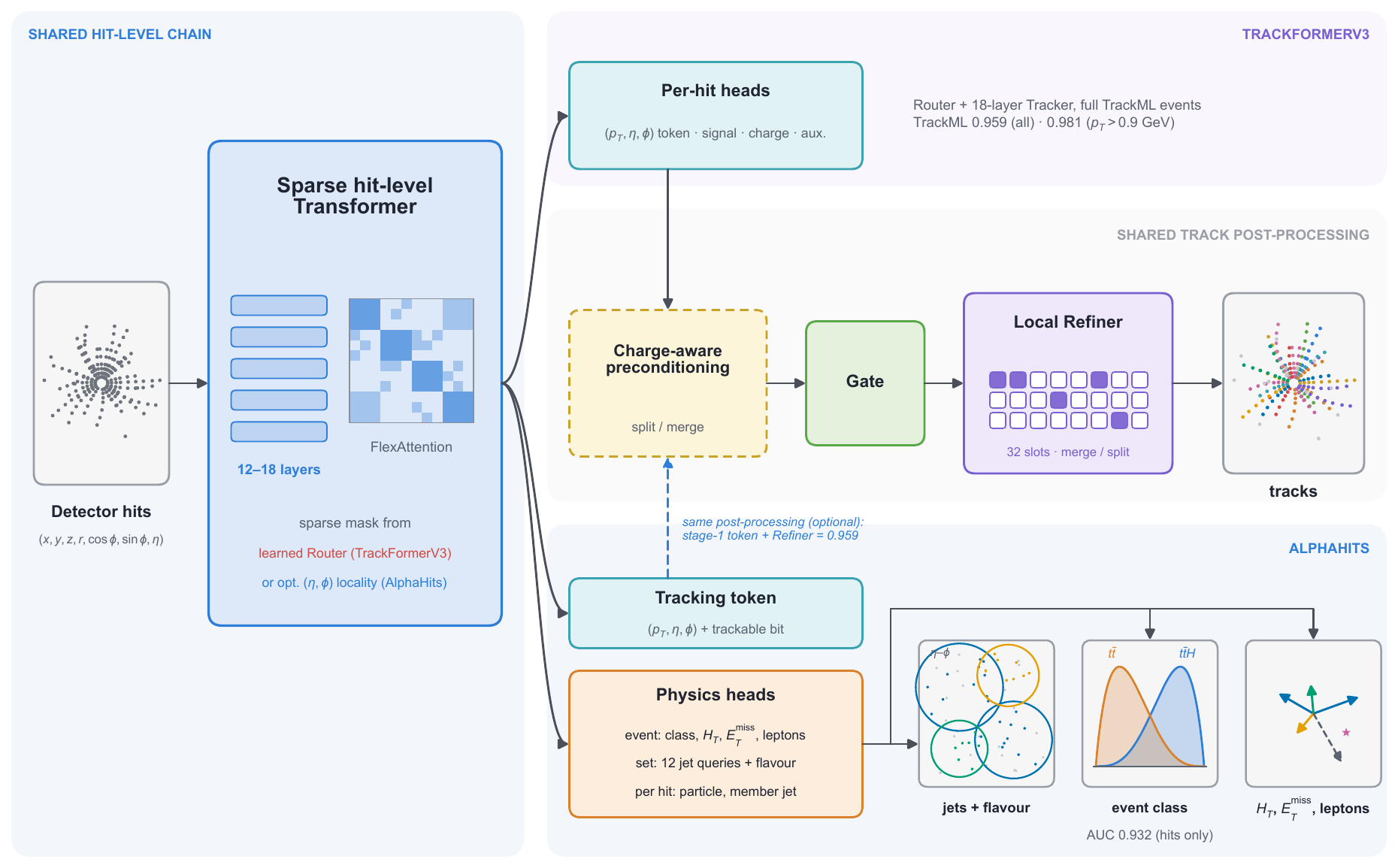}
    \caption{Shared structure of AlphaHits-TrackFormer and AlphaHits. Both models take the detector hits as a set of short feature vectors and pass all hits of an event through a sparse Transformer built on FlexAttention: 18 layers with a learned Router in AlphaHits-TrackFormer, 12 layers with dense attention and an optional $(\eta,\phi)$ locality window or Router in AlphaHits. AlphaHits-TrackFormer (top) predicts for every hit a $(p_{\mathrm{T}},\eta,\phi)$ track token, a signal probability, the charge sign and continuous track parameters, and refines the candidate tracks with a charge-aware preconditioner, a learned Gate and a local Refiner with 32 track slots (middle). AlphaHits (bottom) keeps the tracking token and adds physics heads for jets with flavour, the event class, $H_{\mathrm{T}}$, $E_{\mathrm{T}}^{\mathrm{miss}}$ and leptons. Its tracking output can optionally pass through the same local stage. AlphaHits-TrackFormer additionally uses detector identifiers and cluster-shape features per hit. AlphaHits can, depending on the setup, use calorimeter towers as additional inputs.}
    \label{fig:common_components}
\end{figure}

AlphaHits-TrackFormer and AlphaHits share a common architectural
foundation, as illustrated in Figure~\ref{fig:common_components}.
Both operate directly on collections of inner-tracker hits
and use Transformer encoders to learn representations from
detector-level information. A central development of the
present work is the extension to substantially larger input
multiplicities than in the original studies.
In particular, AlphaHits-TrackFormer processes all pixel
and strip detector hits in complete events containing up to
$1.5\times10^5$ hits, while AlphaHits extends the learned
hit-level representations to multi-task event reconstruction
and classification.

\subsection{Common hit-level representation}
Hits, used as an input, are represented as short vectors of their position and local geometry (Tables~\ref{tab:tf:inputs} and~\ref{tab:hf:inputs}). They are embedded by one linear layer into d=256 vectors, and passed through a stack of Transformer encoder layers. The output of the encoder is one state per hit, and every head of both models reads these per-hit states. The resulting per-hit states form contextualized detector-level representations from which task-specific predictions are obtained. 
AlphaHits-TrackFormer adds learned embeddings of the detector volume, layer and module and eight cluster-shape features for every input hit. AlphaHits adds eight learnable register tokens that are prepended to the hit sequence and stripped before any head.

\subsection{Physics-informed multi-task learning}

A second common feature of the updated models is the use of auxiliary physics objectives. Rather than optimizing only the final task label, the hit representation is simultaneously trained to encode additional reconstruction-level or physics-level quantities.

AlphaHits-TrackFormer predicts, for every hit, a signal probability (i.e., whether the hit was produced by a particle with at least 3 hits in the event), the charge sign and eight continuous track parameters next to the discrete track token. 

AlphaHits predicts, for every hit, its particle’s kinematics, production-vertex displacement and charge, and its member jet’s axis, pT and secondary-vertex flight distance, next to the track token. These auxiliary terms enter the loss with fixed weights outside the main objective (Section~\ref{sec:ah:training}; Appendix~\ref{sec:hf:targets} and Table~\ref{tab:hf:losses} give every term).

\subsection{Sparse full-event attention}
\label{sec:sparse}
Both models use Transformer attention over detector hits, with different attention strategies adapted to their event sizes. AlphaHits-TrackFormer processes $10^5$ hits per event and combines learned sparse attention with FlexAttention to make the computation tractable. In the AlphaHits-TrackFormer pipeline, a learned Router assigns each hit to two of $4 \times 4 \times 32$ groups made in hit-radius, track-pseudorapidity and track-azimuth, with capacity of 768 hits per group. Attention then is only calculated for hit pairs from the same or neighboring groups, avoiding the dense attention costs. 

In this work, AlphaHits operates on hard scatter only events without pile-up, which contain $2\,454$ hits on average and $6\,245$ hits at most. A geometric locality mask, similar to the one used in AlphaHits-TrackFormer, was tested and gave neither computational nor performance differences compared to dense attention due to the small number of hits per event. 

\subsection{From tracker to event model}
As shown in Figure \ref{fig:common_components}, AlphaHits-TrackFormer is a tracker, which predicts the track parameters over all hits, and a local stage chain which resolves ambiguities (Section ~\ref{sec:trackformer}). AlphaHits is the same tracker model with additional heads. Its backbone is first trained on tracking alone, and the nine heads are then trained jointly from the tracking pretrained backbone (Section~\ref{sec:hf:model}). The comparative summary of the models is presented in Table \ref{tab:models}.

\begin{table}[H]
  \centering\small
  \setlength{\tabcolsep}{7pt}\renewcommand{\arraystretch}{1.15}
  \caption{Comparative summary of AlphaHits-TrackFormer and AlphaHits. Inputs, heads and training are described in Sections~\ref{sec:trackformer} and~\ref{sec:hf:model} and in the Appendix.}
  \label{tab:models}
  \begin{tabular}{@{}lll@{}}
    \toprule
     & AlphaHits-TrackFormer & AlphaHits \\
    \midrule
    Events & \begin{tabular}[t]{@{}l@{}}TrackML $t\bar{t}$, $\langle\mu\rangle=200$\\ $1.1\times10^5$ hits per event, up to $1.5\times10^5$\end{tabular}
           & \begin{tabular}[t]{@{}l@{}}$t\bar{t}H$ / $t\bar{t}$, no pile-up\\ 2\,454 hits per event, up to 6\,245\end{tabular} \\
    \addlinespace
    Hit features & 9 position, 3 detector, 8 cluster shape & 7 position \\
    \addlinespace
    Encoder & \begin{tabular}[t]{@{}l@{}}18 layers, $d=256$, 8 heads, FF 1\,536\\ dropout 0.1, pre-norm, no registers\end{tabular}
            & \begin{tabular}[t]{@{}l@{}}12 layers, $d=256$, 8 heads, FF 1\,024\\ dropout 0.2, post-norm, 8 registers\end{tabular} \\
    \addlinespace
    Attention & \begin{tabular}[t]{@{}l@{}}FlexAttention, learned routing\\ ($4\times4\times32$ groups, top-2, capacity 768)\end{tabular}
              & \begin{tabular}[t]{@{}l@{}}FlexAttention, dense with padding mask\\ optional $(\eta,\phi)$ window or learned routing\end{tabular} \\
    \addlinespace
    Track token & $20\times129\times64$ bins & $32\times128\times64$ bins \\
    Other per-hit outputs & signal, charge, 8 track parameters & signal, particle, member jet \\
    Event outputs & -- & class, $H_{\mathrm{T}}$, $E_{\mathrm{T}}^{\mathrm{miss}}$, leptons, $\le12$ jets \\
    Second stage & preconditioner, Gate, local Refiner (32 slots) & nine heads, joint fine-tuning \\
    Parameters & Tracker 19.4\,M; Refiner 3.2\,M; Gate 5\,k & 12.45\,M (9.74\,M pre-trained) \\
    \bottomrule
  \end{tabular}
\end{table}

\section{AlphaHits-TrackFormer}\label{sec:trackformer}

\begin{figure}[H]
    \centering
    \includegraphics[width=\linewidth]{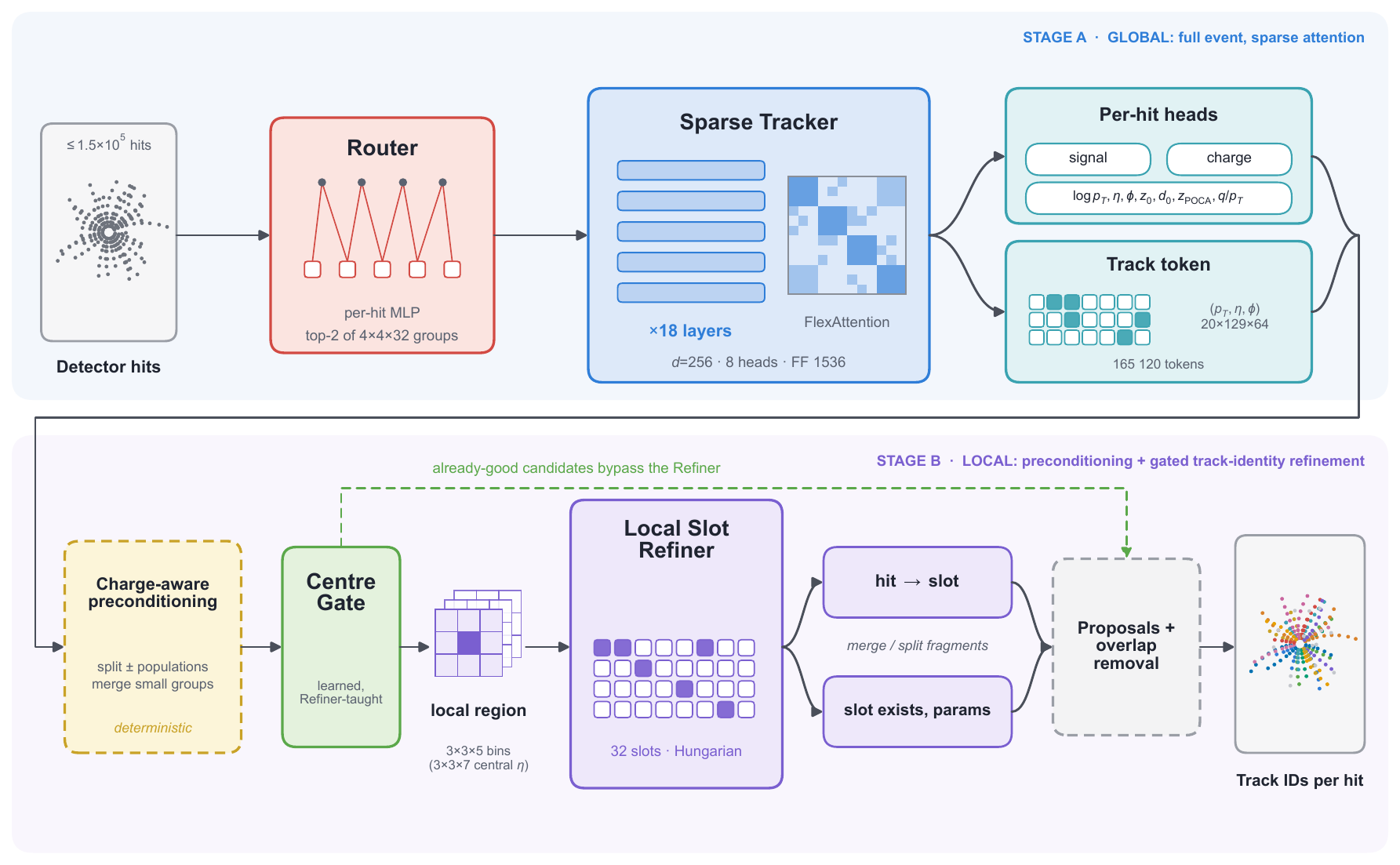}
    \caption{The AlphaHits-TrackFormer pipeline. Stage A, global: a per-hit MLP Router assigns every hit to two of $4\times4\times32$ groups in $(r,\eta,\phi)$, which define the block-sparse FlexAttention of the 18-layer Tracker ($d=256$, 8 heads, feed-forward 1536). For every hit the Tracker predicts a signal probability, the charge sign, eight track parameters and a $(p_{\mathrm{T}},\eta,\phi)$ token of a $20\times129\times64$ codebook. Hits with the same token form a candidate track. Stage B, local: a deterministic charge-aware preconditioner splits and merges candidates, a learned Gate selects the candidate centres that need refinement, and the Refiner reassigns the hits of a $3\times3\times5$-bin neighbourhood ($3\times3\times7$ in the central $\eta$ region) to 32 track slots. Proposals are ranked, overlaps removed, and every hit receives at most one track. Centres not selected by the Gate keep their preconditioned assignment.}
    \label{fig:trackformer_architecture}
\end{figure}

The AlphaHits-TrackFormer reconstruction pipeline is hierarchical. A learned \emph{Router} first assigns detector hits to overlapping logical groups that determine the sparse attention structure. A full-event \emph{Tracker} then predicts track parameters for every hit and maps the hits into a discrete track-parameter space. These predictions are converted into improved preliminary candidates by a lightweight charge-aware preconditioning step. Finally, a learned \emph{Gate} identifies regions that require additional processing, and a compact local \emph{Refiner} resolves the remaining track-identity ambiguities. The architecture of the model is shown in Figure \ref{fig:trackformer_architecture}. In this section we describe the training procedure and every step of the AlphaHits-TrackFormer reconstruction pipeline. 

\subsection{Dataset and input}
AlphaHits-TrackFormer is trained and evaluated on the TrackML dataset~\cite{Amrouche2020TrackML,Kiehn2019TrackMLKaggle}: $t\bar{t}$ events at $\langle\mu\rangle=200$ in a silicon tracker of pixel, short-strip and long-strip layers. An event holds $1.1\times10^5$ hits per event on average and up to $1.5\times10^5$. The dataset offers $8\,700$ events, of which $7\,830$ are used for training, 50 for validation and 820 for testing. The input to the Trackformer consists of 20 features per hit. Twelve of them relate to hit position and geometry. Eight cluster shape features are derived from the digitised cells of the hit and contain: number of cells, total charge, charge fraction of the largest cell, width in cells along each of the two local axes, charge-weighted standard deviation along each axis, and their covariance. Input features are described in Table \ref{tab:tf:inputs}.

\begin{table}[!htb]
  \centering\small
  \setlength{\tabcolsep}{9pt}\renewcommand{\arraystretch}{1.15}
  \caption{Per-hit input features of AlphaHits-TrackFormer. The conformal coordinates make curved trajectories in the transverse plane approximately straight. Cluster features are computed from the digitised cells of the TrackML \texttt{cells} file.}
  \label{tab:tf:inputs}
  \begin{tabular}{@{}lll@{}}
    \toprule
    Group & Feature & Definition \\
    \midrule
    position (9) & $x,\ y,\ z$ & global cluster position / 1000 \\
     & $r$ & $\sqrt{x^2+y^2}$ \\
     & $\sin\varphi,\ \cos\varphi$ & $\varphi=\operatorname{atan2}(y,x)$ \\
     & $\eta$ & $-\ln\tan(\theta/2)$, $\theta=\operatorname{atan2}(r,z)$ \\
     & $u,\ v$ (conformal) & $1000x/r^2$, $1000y/r^2$ \\
    \addlinespace
    detector (3) & volume, layer, module & learned embeddings ($32\times8$, $32\times8$, $4096\times16$) \\
    \addlinespace
    cluster (8) & $n_{\mathrm{cells}}$, $\sum q$ & $\log(1+n_{\mathrm{cells}})$, $\log(1+\sum q)$ \\
     & $\max q/\sum q$ & charge fraction of the largest cell \\
     & $w_0$, $w_1$ & cluster extent along the two local directions \\
     & $\sigma_0$, $\sigma_1$, $\mathrm{cov}_{01}$ & spread and correlation along the two local directions \\
    \bottomrule
  \end{tabular}
\end{table}


\subsection{Router: learned group assignment}\label{sec:tf:router}
The Router is a small MLP that processes hits independently. It receives the continuous hit features together with learned embeddings describing the detector volume, layer, and module, and predicts membership in a $4\times4\times32$ grid in radial position, coarse pseudorapidity, and periodic azimuth. This corresponds to 512 logical routing groups. Each hit is assigned to its two most probable routes.

Routed hit copies are packed into groups of capacity 768, giving a maximum packed sequence length of $393\,216$ routed entries. This number describes the capacity of the routed representation rather than the number of physical detector hits, since each physical hit can appear in two routing groups. Primary routes are packed first, so that secondary copies are preferentially removed if a group overflows.

The subsequent attention calculation in the next pipeline stage is restricted to hits within the same routing group and selected neighboring groups. The Router therefore serves two purposes: it makes full-event attention computationally tractable and introduces a learned locality structure determining which detector regions can communicate directly.

\subsection{Tracker: full-event track-parameter prediction}
An 18-layer Transformer encoder is run over the routed event and predicts for every hit a discrete track token: independent $p_{\mathrm{T}}$, $\eta$, and $\phi$ bins of a $20\times129\times64$ codebook ($165\,120$ tokens in total). This binning strategy allows 100.000 of tokens and we found a clear outperformance compared to VQ-VAE based approaches, see also Appendix \ref{sec:binning}.
Additionally, it outputs a signal probability (hit belonging to a particle with $\geq3$ hits), the charge sign and 8 continuous parameters: ($\log\pt$, $\eta$, $\sin\phi$, $\cos\phi$, $z_0$, $d_0$, $z_{\mathrm{POCA}}$, $q/\pt$). Hits with the same $(\pt,\eta,\phi)$ token form a candidate track. The token separates global localisation from exact identity. Hits originating from the same charged particle are expected to populate the same or nearby regions of the track-parameter codebook. Small per-hit classification differences can nevertheless fragment one physical trajectory across neighboring candidate tokens, while multiple physical tracks can occasionally be assigned to the same token. These residual ambiguities motivate the subsequent local reconstruction stages.





\subsection{Charge-aware candidate preconditioning}
A deterministic step splits candidate groups whose hits carry confident predictions of both charge signs, and merges nearly empty candidates into a sign-compatible, better-populated neighbour. This operation is computationally inexpensive and improves the preliminary reconstruction before the learned Refiner is applied. Importantly, it is used as a preprocessing step rather than as a hard statement about track identity: the subsequent learned model retains access to the relevant Tracker and preconditioning information and can modify these assignments when necessary.




\subsection{Learned gating of local refinement}
Although local refinement substantially improves reconstruction quality, running the Refiner around every populated candidate is computationally unnecessary. We therefore introduce a lightweight learned Gate that predicts whether a candidate region requires local refinement.

The Gate is an MLP that uses inexpensive information already available from the Tracker and preconditioning stages, including candidate occupancies, neighboring-group structure, signal and charge information, and properties of the predicted track-parameter distributions. It is trained using the full Refiner as a teacher: candidate centres are labeled according to whether their Refiner proposals contribute to the final reconstructed hit assignment.

At inference time, only candidate regions passing the Gate are sent to the local Transformer. The gate threshold therefore provides a direct trade-off between reconstruction accuracy and computational cost. In the operating region studied here, a substantial fraction of local Refiner evaluations can be removed with only a very small reduction in the final TrackML score.


\subsection{Refiner: local track-identity reconstruction}\label{sec:tf:refiner}

The remaining combinatorial ambiguities are resolved by a compact local Transformer. For each selected candidate centre, the Refiner collects hits from a neighborhood in predicted $(p_{\mathrm{T}},\eta,\phi)$ space. The standard neighborhood spans $3\times3\times5$ bins in $p_{\mathrm{T}}$, $\eta$, and $\phi$. In the central pseudorapidity region, where the global reconstruction is more challenging, the azimuthal context is widened to give a $3\times3\times7$ neighborhood.

Rather than assigning hits directly to globally labelled candidate classes, the Refiner assigns the hits from one neighborhood to a fixed number of local track slots. The slots are not predefined, and the model decides which of them to use for the tracks present in each neighborhood, so some slots can represent tracks while others remain unused. The Refiner can therefore perform two complementary operations: \textbf{merge} neighboring Tracker fragments when they originate from the same physical trajectory, or \textbf{split} a coarse Tracker candidate when several physical trajectories have been assigned to the same global token.

The Refiner is trained using the global Tracker output together with information from the preconditioning stage. During training, Hungarian matching is used to associate the track slots with truth particles in a minimum-cost one-to-one mapping. The objective of the Refiner includes hit-to-slot assignment, slot-existence and track-parameter terms, together with track-weight-aware supervision and a soft track-level purity and coverage objective. In this way, the local model is trained specifically to correct the residual failures of the global reconstruction.

In the final output of the Refiner, neighborhoods can overlap and one hit can appear in proposals of several candidate tracks. Therefore, in one last post-processing step, proposals are ranked and pruned by non-maximum suppression, and every hit is assigned to at most one track.

\subsection{Results}
Table \ref{tab:tf:ladder} shows the performance of the same 18-layer Tracker, and each stage of the AlphaHits-TrackFormer pipeline on the 820 test events. The Tracker alone reaches 0.844 over all particles and 0.912 for $\pt>0.9$\,GeV. The full pipeline reaches 0.959 and 0.981. The gain is local: the Tracker's token is exactly right for 82-87\,\% of the hits per axis but within one bin for 93-98\,\% (Table~\ref{tab:tf:ladder}, lower block), so the Refiner mostly has to merge neighbours.

\begin{figure}[H]
    \centering
    \includegraphics[scale=0.6]{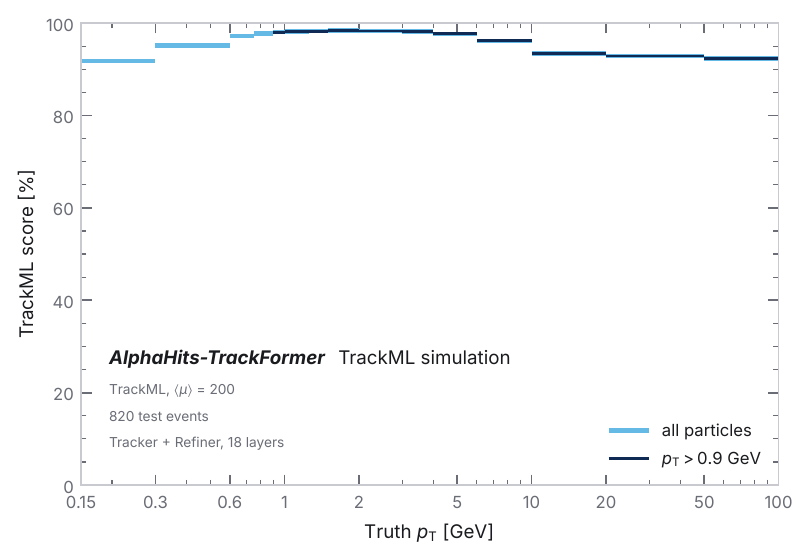}
    \includegraphics[scale=0.6]{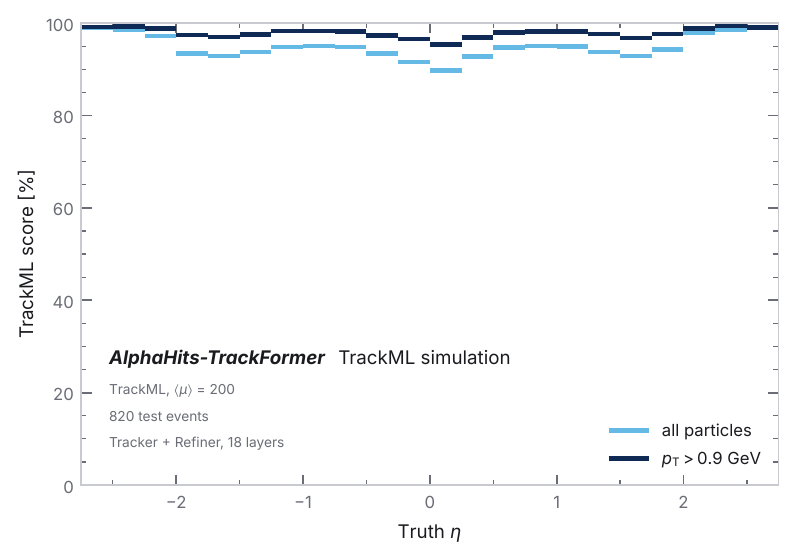}
    \caption{TrackML score of AlphaHits-TrackFormer (18-layer Tracker, preconditioner and Refiner, no Gate) on the 820 test events, per bin of true particle $p_{\mathrm{T}}$ (left) and $\eta$ (right), for all particles and for particles with $p_{\mathrm{T}}>0.9$\,GeV. The last $p_{\mathrm{T}}$ bin collects all particles above 50\,GeV, and the outer $\eta$ bins all particles with $|\eta|>2.5$.}
    \label{fig:differential_pt}
\end{figure}

\begin{table}[H]
  \centering\small
  \setlength{\tabcolsep}{6pt}\renewcommand{\arraystretch}{1.15}
  \caption{AlphaHits-TrackFormer with the 18-layer Tracker, stage by stage. Upper block: TrackML score on the 820 test events, over all particles, for $p_{\mathrm{T}}>0.9$\,GeV and in slices of $|\eta|$ (for $p_{\mathrm{T}}>0.9$\,GeV). The fraction of candidate centres sent to the Refiner and the mean time per event (Router and Tracker plus the local stage) are measured on 50 of the events on an NVIDIA H100, with an inference batch of 256 neighbourhoods. The scan over Gate thresholds is in Table~\ref{tab:tf:ladder_full}. Lower block: per-hit accuracies of the Tracker alone on the same 820 events.}
  \label{tab:tf:ladder}
  \begin{tabular}{@{}lcccccccc@{}}
    \toprule
    Stage & all & $p_{\mathrm{T}}>0.9$ & $|\eta|<1$ & $1\le|\eta|<2$ & $2\le|\eta|<3$ & $|\eta|\ge3$ & centres refined & time [s] \\
    \midrule
    Tracker & 0.844 & 0.912 & -- & -- & -- & -- & 0\,\% & 0.45 \\
    + preconditioner & 0.880 & 0.934 & 0.923 & 0.927 & 0.955 & 0.935 & 0\,\% & 0.47 \\
    + Refiner, no Gate & 0.959 & 0.981 & 0.973 & 0.975 & 0.992 & 0.989 & 100\,\% & 4.60 \\
    + Refiner, Gate 0.30 & 0.959 & 0.980 & 0.972 & 0.974 & 0.991 & 0.989 & 56\,\% & 2.98 \\
    \bottomrule
  \end{tabular}

  \vspace{8pt}
  \begin{tabular}{@{}lccc@{}}
    \toprule
    Tracker alone, per hit & $p_{\mathrm{T}}$ & $\eta$ & $\phi$ \\
    \midrule
    token bin exact & 82.6\,\% & 84.3\,\% & 87.3\,\% \\
    token bin within $\pm1$ & 97.7\,\% & 94.2\,\% & 93.3\,\% \\
    \addlinespace
    charge sign & \multicolumn{3}{c}{97.1\,\%} \\
    signal (trackable) hit & \multicolumn{3}{c}{95.2\,\%} \\
    \bottomrule
  \end{tabular}
\end{table}

A breakdown of the results of the best AlphaHits-TrackFormer model (18L Tracker + Refiner) for different $\eta$ and $\pt$ bins is shown in Figures~\ref{fig:differential_pt} and \ref{fig:differential_eta}. These results highlight strong performance at lower $\pt$ values and across almost the full $\eta$ range. Above 10 GeV the score falls to about 0.93. Such particles are rare, about 6 of the 9,200 particles in an event, so the training sample holds few of them.

The charge sign, which is not part of the token, is right for 97.1\% of the hits, so the global representation carries the curvature sign. The signal probability separates hits of trackable particles from the rest with 95.2\% accuracy.

\section{AlphaHits}\label{sec:hf:model}
\begin{figure}[!htb]
    \centering
    \includegraphics[width=\linewidth]{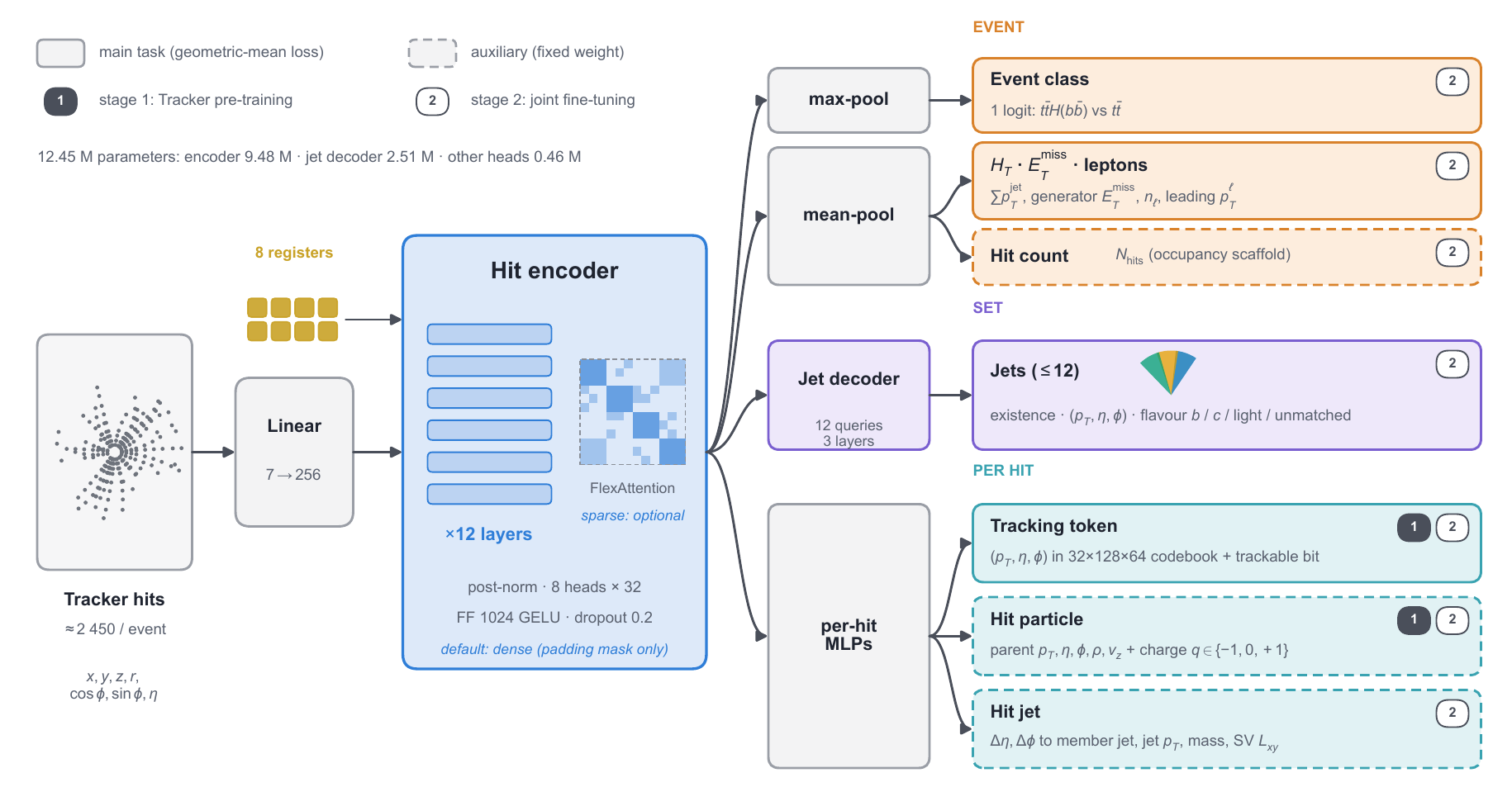}
    \caption{The AlphaHits model. Each hit enters with seven position features, is embedded by one linear layer into $d=256$, and passes with eight register tokens through a 12-layer encoder (post-norm, 8 heads, feed-forward 1024, dropout 0.2; dense attention with a padding mask, optionally sparse). Three readouts feed nine heads: max-pooled hit states for the event class, mean-pooled states for $H_{\mathrm{T}}$, $E_{\mathrm{T}}^{\mathrm{miss}}$, leptons and the hit count, a 12-query decoder for the jets with flavour, and per-hit MLPs for the tracking token, the hit's particle and its member jet. Solid boxes are main tasks of the geometric-mean loss and dashed boxes auxiliary terms with fixed weights. The badges (numbers 1 and 2) give the training stage (1: tracking pre-training, 2: joint fine-tuning).}
    \label{fig:hitformer_architecture}
\end{figure}

AlphaHits is a permutation-equivariant Transformer architecture operating directly on detector hits, building on the TrackFormer and HiggsFormer models. The first HiggsFormer was an event classifier over raw hit coordinates $(x,y,z)$ with a single $t\bar{t}H$  ($H\to b\bar b$) vs $t\bar{t}$ head (plus an auxiliary $H_T$ regression head for the `big' model), trained on a 40k-event sample~\cite{Caron2025Higgsformer}. AlphaHits builds on the tracking backbone of Section~\ref{sec:trackformer}. It is pretrained on tracking tasks only (Stage 1), and fine-tuned with nine heads to predict tracks, jets with flavour, $H_{\mathrm{T}}$, $E_{\mathrm{T}}^{\mathrm{miss}}$, leptons and the event class in one forward pass (Stage 2). The architecture is visualised in Figure \ref{fig:hitformer_architecture}.

\subsection{Simulation and dataset}
We simulate $200\,000$ $pp$ events at $\sqrt{s}=14$\,TeV, no pileup, half $t\bar{t}H$ with $H\to b\bar b$ and half $t\bar{t}$. Events are generated with \textsc{Pythia}~8.311 inside ACTS~41.1 with a parton-level cut $\hat p_{\mathrm{T}}>100$\,GeV on the hard process, inclusive top and $W$ decays and no pile-up, and simulated with the ACTS \textsc{Fatras} fast simulation of the silicon-only GenericDetector (the TrackML layout in a 2\,T field). Hits are digitised cluster centroids. The same generator record passes through \textsc{Delphes}~3.5.1 with an ATLAS-like card, which supplies the targets ( GenJets \footnote{anti-$k_t$ $R=0.6$ jets with $p_{\mathrm{T}}>20$\,GeV clustered from all stable generator particles except neutrinos}, $E_{\mathrm{T}}^{\mathrm{miss}}$, secondary vertices, reconstructed leptons) and the two references: the full chain (calorimeters and tracks) and a tracker-only chain. For the calorimeter variant, the Delphes calorimeter towers of the same events are stored as a second input. Each ECal or HCal tower carries the seven position features of its cell centre on the calorimeter surface, its log energy, an ECal/HCal flag and its cell size in $\eta$ and $\phi$ (11 features) of the same events are stored as a second input. Events are split per process with a fixed seed into $160\,000$ / $20\,000$ / $20\,000$ for training, validation and test. All the metrics are reported on the $20\,000$ test events, and every Delphes reference is evaluated on the exact same events. Table \ref{tab:ah:sample} summarises the sample.

\subsection{Inputs}
Each hit is the seven-dimensional vector: position in metres, transverse radius, azimuth as $(\cos\varphi,\sin\varphi)$ and pseudorapidity. The model sees positions only: no detector identifiers, cluster shape, charge or timing. Events are zero-padded within a batch and the padding is masked. Full description of the inputs is presented in the Table \ref{tab:hf:inputs}.

\begin{table}[htbp]
  \centering
  \caption{Per-hit input features of AlphaHits. Positions are the digitised
  cluster centroids in global coordinates, stored in mm and scaled to metres.}
  \label{tab:hf:inputs}
  {\footnotesize\setlength{\tabcolsep}{5pt}\renewcommand{\arraystretch}{1.12}
  \begin{tabular}{@{}llll@{}}
    \toprule
    Feature & Definition & Unit / range \\
    \midrule
    $x,\ y,\ z$ & global cluster position $/\,1000$ & m; $|x|,|y|\le1.03$, $|z|\le2.96$ \\
    $r$ & $\sqrt{x^2+y^2}$ & m; $0.03\le r\le1.03$ \\
    $\cos\varphi,\ \sin\varphi$ & $\varphi=\operatorname{atan2}(y,x)$ & $[-1,1]$ \\
    $\eta$ & $-\ln\tan(\theta/2)$, $\theta=\operatorname{atan2}(r,z)$ & $|\eta|\le4.6$ \\
    \bottomrule
  \end{tabular}}
\end{table}

Delphes calorimeter towers of the same events (261 per event on average, 11 features each) are used as an additional input for the AlphaHits model setup trained on hits and calorimeter towers.

\subsection{Task heads}
The nine heads of the AlphaHits model are listed in Table~\ref{tab:ah:heads}. Three encoder readouts feed the heads: max-pooled hit states for the class logit, mean-pooled states for the scalar heads ($H_T$, $E_T^{miss}$, lepton count and leading-lepton $\pt$, hit count), and the per-hit states for the tracking, hit-particle and hit-jet heads and for the 12-query jet decoder. The tracking head assigns every hit to a $32\times128\times64$ $(\pt,\eta,\phi)$ codebook and predicts a signal bit. Hits with the same token form one track.

\begin{table}[!htb]
  \centering\small
  \setlength{\tabcolsep}{9pt}\renewcommand{\arraystretch}{1.15}
  \caption{AlphaHits heads. Readout: max- or mean-pool over the hit states (event), learnable queries with cross-attention (set), or a token-wise MLP (per hit). Stage: 1 = tracker pre-training, 2 = joint fine-tuning. Auxiliary (aux) heads enter the loss with a fixed weight outside the main combiner. Regressions act on standardised targets; $\phi$ is predicted as $(\sin\phi,\cos\phi)$; the jet flavour classes are $b$, $c$, light and unmatched.}
  \label{tab:ah:heads}
  \setlength{\tabcolsep}{7pt}
  \begin{tabular}{@{}lllrc@{}}
    \toprule
    Head & Readout & Output & Params & Stage \\
    \midrule
    Event class & event, max-pool & 1 logit, $t\bar{t}H$ vs $t\bar{t}$ & 257 & 2 \\
    $H_{\mathrm{T}}$ & event, mean-pool & $\sum_{\mathrm{jets}}p_{\mathrm{T}}$ & 33\,k & 2 \\
    $E_{\mathrm{T}}^{\mathrm{miss}}$ & event, mean-pool & generator-level $E_{\mathrm{T}}^{\mathrm{miss}}$ & 33\,k & 2 \\
    Leptons & event, mean-pool & count, leading-lepton $p_{\mathrm{T}}$ & 33\,k & 2 \\
    Hit count (aux) & event, mean-pool & number of hits & 33\,k & 2 \\
    Jets & set, 12 queries, 3 layers & existence, $(p_{\mathrm{T}},\eta,\sin\phi,\cos\phi)$, 4-class flavour & 2.51\,M & 2 \\
    Tracking & per hit & token bins $32+128+64$, signal bit & 124\,k & 1, 2 \\
    Hit particle (aux) & per hit & $(p_{\mathrm{T}},\eta,\sin\phi,\cos\phi,\rho,v_z)$, 3-class charge & 134\,k & 1, 2 \\
    Hit jet (aux) & per hit & $(\Delta\eta,\Delta\phi)$ to the jet axis, jet $p_{\mathrm{T}}$, mass, SV $L_{xy}$ & 67\,k & 2 \\
    \bottomrule
  \end{tabular}
\end{table}

\subsection{Two-stage training and loss}
\label{sec:ah:training}
Each hit enters as seven features, $(x,y,z,r,\cos\phi,\sin\phi,\eta)$, embedded by one linear layer into $d=256$. Eight learnable register tokens are prepended, and the sequence passes through $L=12$ Transformer encoder layers (8 heads of dimension 32, GELU feed-forward of width 1024, dropout 0.2). When run on both hits and calorimeter towers configuration, AlphaHits adds the Delphes calorimeter towers of the event: 261 per event on average, 11 features each  through zero-initialised, gated cross-attention adapters after encoder layers 6 and 12, so that at step~0 the model is exactly the hits-only model.

 Three readouts feed the heads of Table~\ref{tab:ah:heads}: max-pooled hit states for the class logit, mean-pooled states for the scalar heads, and per-hit states for the dense heads and the 12-query jet decoder. The tracking head assigns every hit to a factorised $(\pt,\eta,\phi)$ codebook of $32\times128\times64=262\,144$ tokens and predicts a trackable bit. 

Stage~1 trains the backbone as a Tracker from scratch. At this stage only the tracking and hit-particle heads exist. Stage~2 adds the other seven heads with fresh weights and trains them jointly, warm-started from the Stage 1 checkpoint (Table~\ref{tab:ah:training}). The Stage 2 loss is a geometric mean over the six main tasks plus fixed-weight auxiliary terms,
\begin{align}
  \label{eq:hf:total1}
  \mathcal L^{(1)} &= L_{\mathrm{trk}} + 0.4\,L_{\mathrm{hitpart}},\\
  \label{eq:hf:total2}
  \mathcal L^{(2)} &= \big(L_{\mathrm{cls}}\,L_{\mathrm{HT}}\,L_{\mathrm{MET}}\,L_{\mathrm{lep}}\,L_{\mathrm{jets}}\,L_{\mathrm{trk}}\big)^{1/6}
     + 0.2\,L_{\mathrm{nhits}} + 0.4\,L_{\mathrm{hitpart}} + 0.3\,L_{\mathrm{hitjet}}.
\end{align}
The geometric mean is scale-free, so the main tasks need no weights and a task with a small loss receives the larger relative gradient. The auxiliary terms stay outside the product so that a near-zero scaffold loss cannot collapse it. Regressions are Huber losses on standardised targets, per-hit terms are masked means, jets are a set loss with Hungarian matching. Appendix~\ref{sec:hf:targets} explains every term.

\subsection{Results}
\label{sec:ah:results}

\begin{figure}[H]
  \centering
  \includegraphics[width=\textwidth]{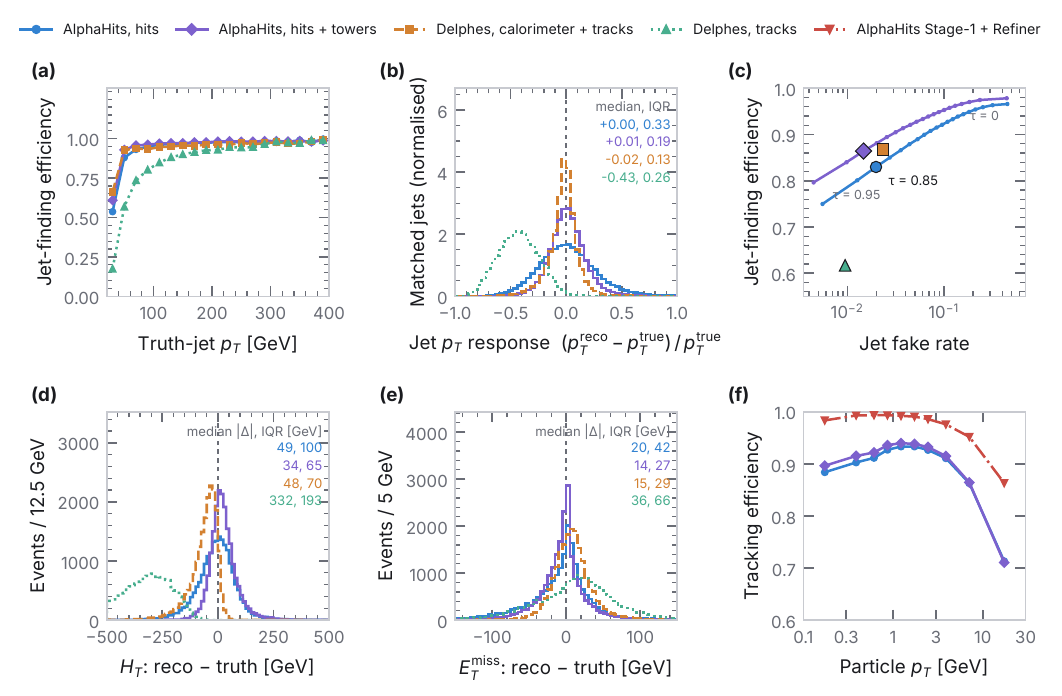}
  \caption{AlphaHits reconstruction on the 20k test events, hits only and hits + calorimeter towers, against the Delphes full chain (calorimeter + tracks) and tracker-only chain on the same events. (a)~Jet-finding efficiency against truth-jet $\pt$ (GenJets, anti-$k_t$ $R=0.6$, $\pt>20$\,GeV; match $\Delta R<0.4$); (b)~relative $\pt$ response of matched jets; (c)~efficiency against fake rate while sweeping the existence threshold, with the operating point 0.85 marked and the Delphes chains as single points; (d,\,e)~residuals of the $H_T$ and $E_T^{miss}$ regressions with median absolute residual and inter-quartile range; (f)~double-majority tracking efficiency against particle $\pt$: two one-shot multitask AlphaHits models with the equal-token decode (blue and purple), and AlphaHits Stage 1 backbone (trained on tracking alone) with the local Refiner.}
  \label{fig:ah:performance}
\end{figure}

Figure~\ref{fig:ah:performance} and Table~\ref{tab:ah:reco} give the reconstruction performance of AlphaHits with hits and 
hits+calorimeter towers as inputs, against the two Delphes chains and evaluated on the same events.

From tracker hits alone, AlphaHits uses comparable detector information to the Delphes tracker-only reconstruction chain. AlphaHits finds 83.0\,\% of the GenJets at a 2.0\,\% fake rate (existence threshold 0.85), against 61.7\,\% at 0.9\,\% for the track jets of the tracker-only Delphes chain. The AlphaHits jet $p_{\mathrm{T}}$ is unbiased (median response $+0.003$), whereas the track jets contain only charged particles and are 43\,\% too low. The event-level quantities follow the same pattern. The $H_T$ and $E_T^{\mathrm{miss}}$ that AlphaHits predicts from the hits differ from the true values by 49 and 20\,GeV in the median. The tracker-only chain misses the neutral energy: its $H_T$ is 332\,GeV too low in the median, and its $E_T^{\mathrm{miss}}$ is off by 36\,GeV (Figure~\ref{fig:ah:performance}, Table~\ref{tab:ah:reco}).

With the calorimeter towers as an additional input, AlphaHits uses comparable detector information to the Delphes full chain of calorimeter and tracks. AlphaHits finds 86.4\,\% of the GenJets at a 1.5\,\% fake rate, against 86.7\,\% at 2.3\,\% for the full Delphes chain. Its jet $p_{\mathrm{T}}$ is unbiased (median response $+0.008$, against Delphes' $-0.017$) but less precise, with an inter-quartile range of 0.19 against 0.13. AlphaHits' $H_T$ and $E_T^{\mathrm{miss}}$ differ from the true values by 34 and 14\,GeV in the median, against 48 and 15\,GeV for the full chain. The towers also improve the flavour rejection for AlphaHits (Figure~\ref{fig:ah:ftag}) and slightly improve the tracking score.



\begin{table}[H]
  \centering\small
  \setlength{\tabcolsep}{9pt}\renewcommand{\arraystretch}{1.15}
  \caption{Reconstruction on the 20k test events: the numbers behind Figure~\ref{fig:ah:performance}. Models at existence threshold 0.85; jet response $=(p_{\mathrm{T}}^{\mathrm{reco}}-p_{\mathrm{T}}^{\mathrm{true}})/p_{\mathrm{T}}^{\mathrm{true}}$ for matched jets; residuals are quoted as median absolute residual / inter-quartile range. Tracking: TrackML score with the equal-token decode (codebook ceiling 0.959) and per-particle double-majority efficiency for $p_{\mathrm{T}}>1$\,GeV.}
  \label{tab:ah:reco}
  \begin{tabular}{@{}lcccc@{}}
    \toprule
     & \multicolumn{2}{c}{AlphaHits} & \multicolumn{2}{c}{Delphes} \\
    \cmidrule(lr){2-3}\cmidrule(lr){4-5}
     & hits & hits + towers & calo + tracks & tracks \\
    \midrule
    Jet-finding efficiency & 0.830 & 0.864 & 0.867 & 0.617 \\
    Jet fake rate & 0.020 & 0.015 & 0.023 & 0.009 \\
    Jet $p_{\mathrm{T}}$ response, median / IQR & $+0.003$ / 0.33 & $+0.008$ / 0.19 & $-0.017$ / 0.13 & $-0.43$ / 0.26 \\
    $H_{\mathrm{T}}$ residual [GeV], median / IQR & 49.4 / 99.9 & 33.9 / 65.5 & 48.1 / 70.2 & 332.5 / 192.5 \\
    $E_{\mathrm{T}}^{\mathrm{miss}}$ residual [GeV], median / IQR & 19.9 / 41.7 & 13.8 / 26.9 & 15.3 / 29.1 & 36.2 / 66.3 \\
    Tracking score, equal-token decode & 0.806 & 0.819 & -- & -- \\
    $\varepsilon_{\mathrm{DM}}$ ($p_{\mathrm{T}}>1$\,GeV) & 0.893 & 0.897 & -- & -- \\
    \bottomrule
  \end{tabular}
\end{table}

The one-shot model keeps the tracking head, but tracking is now one of nine tasks sharing the
encoder. On the test events it reaches a TrackML score of 0.806 with the equal-token decode and a double-majority efficiency of 0.893 for $\pt>1$\,GeV. The towers improve the tracking (0.819, 0.897). However, the fair measure of the backbone as a tracker is the Stage 1 model, the same network before the heads are added: trained on tracking alone it reaches 0.891 (efficiency 0.929), 0.927 with the validation-tuned decode and 0.959 (100\% of the ceiling) with a local Refiner of the AlphaHits-TrackFormer type trained on its output (Table~\ref{tab:ah:tracking}). Note that the AlphaHits backbone is smaller than the AlphaHits-TrackFormer Tracker (12 instead of 18 layers, 9.7\,M instead of 19.4\,M parameters) and is trained on a different sample with a different codebook, so the two tracking scores are not directly comparable. The depth dependence of Figure~\ref{fig:tracking-scaling} suggests that a deeper backbone would also lead to a better tracking performance.

\subsection{Jet flavour classification} 
\begin{figure}[H]
    \centering
    \includegraphics[scale=0.9]{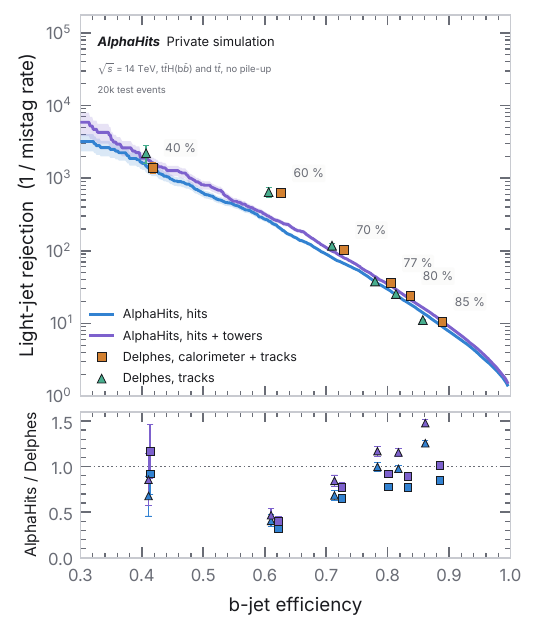}
    \includegraphics[scale=0.9]{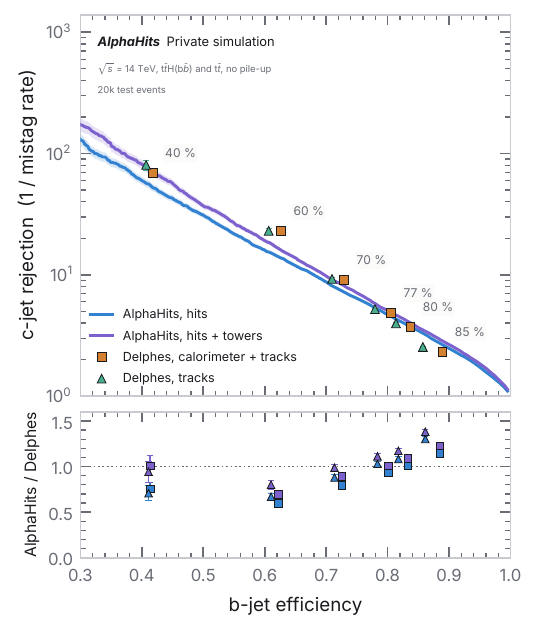}
    \caption{$b$-tagging with the jet decoder's $P(b)$ on the 20k test events: (left)~light-jet rejection and (right)~$c$-jet rejection against $b$-jet efficiency, for hits only and hits + calorimeter towers, with the Delphes calorimeter-jet and track-jet taggers at measured $b$-efficiencies of 40--85\,\%. Delphes tags are truth-seeded, while the model infers flavour from the hits.}
    \label{fig:ah:ftag}
\end{figure}

Figure~\ref{fig:ah:ftag} shows the $b$-tagging performance of the jet decoder's $P(b)$ on the 20k test events: light-jet and $c$-jet rejection against $b$-jet efficiency, for hits only and for hits + calorimeter towers. The reference points are the Delphes calorimeter-jet and track-jet taggers at measured $b$-efficiencies of 40, 60, 70, 77, 80 and 85\,\%. Delphes tags are truth-seeded: each jet is tagged with the card's $\pt$-dependent probability for its true flavour (Appendix~\ref{app:ah:btag}), so the points are an idealised reference. Every jet is graded against the same truth: a reconstructed jet takes the flavour of the nearest GenJet within $\Delta R<0.4$, and the light background is light quarks and gluons.

From hits alone, the model matches the Delphes track-jet reference from 77\,\% $b$-efficiency on (light-jet rejection ratio 1.0, 1.0 and 1.3 at 77, 80 and 85\,\%; $c$-jet 1.0, 1.1 and 1.3) and falls below it at tighter working points (0.68 and 0.88 at 70\,\%; 0.41 and 0.67 at 60\,\%). Against the calorimeter-jet reference it reaches 0.77--0.85 of the light-jet rejection and 0.94--1.14 of the $c$-jet rejection from 77\,\% on. The towers raise every rejection by 10 to 35\,\%. From 77\,\% on, the towers model rejects more light and $c$ jets than the track-jet reference (1.2--1.5) and as many as the calorimeter-jet reference (0.9--1.0 light, 1.0--1.2 $c$). In absolute terms, at 77\,\% $b$-efficiency the hits-only model keeps one light jet in 38 and one $c$ jet in 5, from hit positions alone, with no intermediate separate track reconstruction or vertexing.

\subsection{Higgs event classification}

\begin{figure}[H]
    \centering
  \includegraphics[width=0.8\textwidth]{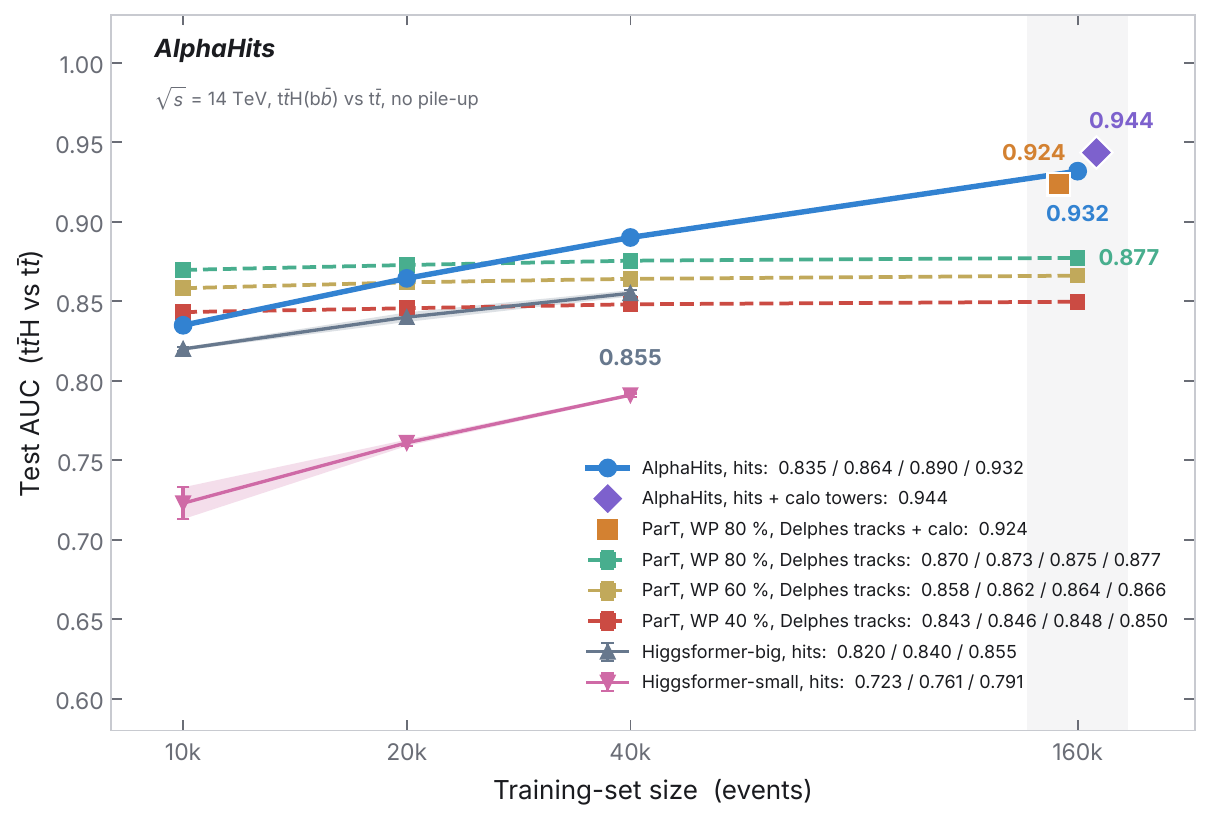}
  \caption{Test AUC against training-set size on the same 20k test events except the 2025 HiggsFormer. AlphaHits retrained with both stages on hits at 10k, 20k, 40k and 160k events, and with calorimeter towers at 160k. ParT on Delphes objects at measured $b$-tagging efficiencies of 80, 60 and 40\,\%. Delphes tracks: track jets (anti-$k_t$, $R=0.6$, $p_T>20$\,GeV) with $p_T$, $\eta$, $\phi$, $E$ and a binary $b$-tag, and the track $E_T^{miss}$. Delphes tracks + calo (160k, 80\,\%): calorimeter jets with the same features and $b$-tag, the calorimeter $E_T^{miss}$, and electrons, muons removed.}
  \label{fig:ah:scaling}
\end{figure}

Figure~\ref{fig:ah:scaling} extends the learning-curve study
of Ref.~\cite{Caron2025Higgsformer}. Using tracker hits alone,
AlphaHits achieves an AUC of 0.932, outperforming the original
HiggsFormer at the same training-set size. In comparison,
ParT using Delphes track jets and track $E_T^{miss}$ with an 80\,\% $b$-tagging
working point achieves an AUC of 0.877, while ParT using
Delphes calorimeter jets, missing transverse momentum,
and electrons reaches 0.924 with 160\,000 training events.
Including calorimeter towers as additional inputs increases
the AlphaHits AUC to 0.944 on the same test events.

\section{Discussion}

The AlphaHits-TrackFormer results demonstrate that
Transformer-based charged-particle tracking can scale to
complete high-occupancy HL-LHC events, processing both pixel and
strip detector hits without restricting reconstruction to
high-$p_T$ particles. The architecture combines two
complementary stages: global sparse attention provides
event-wide context and maps hits into a physically meaningful
track-parameter representation, while local learned refinement
resolves the combinatorial ambiguities required for track
assembly. The strong dependence on network depth (see also Figure~\ref{fig:tracking-scaling}) observed
in high-statistics training further highlights the importance
of model capacity and sufficient training data, in contrast
to earlier low-statistics studies where deeper models
provided limited benefits.

AlphaHits extends this approach beyond tracking through
tracking-based pretraining followed by joint multi-task
reconstruction and classification. The results demonstrate
that a shared hit-level representation can support track
and jet reconstruction, jet flavour classification,
event-level observable prediction, and Higgs event
classification. In particular, AlphaHits achieves an
AUC of 0.932 using tracker hits alone, to be compared with the
0.924 obtained with a conventional tracker-plus-calorimeter
reconstruction chain using Particle Transformer. Adding calorimeter towers to the AlphaHits inputs increases
the AUC to 0.944, re-demonstrating performance comparable
to the full reconstruction baseline.

These developments represent steps towards a broader
vision of hit-level foundation models for LHC physics.
Rather than relying on independent algorithms for each
stage of reconstruction, such models could learn reusable
representations directly from detector measurements,
supporting multiple reconstruction and analysis tasks
within a common framework. Beyond the tasks studied
here, these representations could potentially be adapted
to vertex reconstruction, particle identification,
anomaly detection, and searches for new physics.
More broadly, the objective is to optimize the analysis
chain end-to-end for specific physics goals, potentially
surpassing the performance of conventional reconstruction
pipelines and enabling sensitivity to signals that are
currently difficult or impossible to reconstruct.

Several limitations and opportunities remain. While
AlphaHits-TrackFormer demonstrates full-event tracking
at HL-LHC-like occupancies, the multi-task AlphaHits
model is currently evaluated on hard-scatter events
without pileup. Extending joint reconstruction and
classification to complete high-pileup events remains
an important next step. Furthermore, tracking-based
pretraining provides an initial mechanism for learning
reusable representations, but systematic studies of
transfer learning, self-supervised pretraining, and
generalization across physics processes and detector
configurations are needed to establish broader
foundation-model capabilities. Detailed ablations of
the auxiliary objectives, together with systematic
evaluations of computational efficiency and reconstruction
performance across detector regions and particle
kinematics, will further clarify the strengths and
limitations of the approach.
Moreover, the learning curves show that the present
models remain strongly limited by training-data statistics,
indicating substantial potential for further improvements
through larger-scale training. Lastly, the focus of this work was the development of hit-level tracking and foundation models and their physics performance, so the optimisation of inference time latency is left for future efforts. 

%
\section{Conclusion}

We have introduced AlphaHits, a Transformer-based approach towards hit-level foundation models for LHC physics, building on the TrackFormer and HiggsFormer architectures. AlphaHits-TrackFormer demonstrates scalable full-event charged-particle tracking, processing both pixel and strip detector hits in events containing up to \(1.5\times10^5\) hits.

A central result of this work is that the reconstruction is performed inclusively over the full reconstructable charged-particle population represented in the TrackML sample, without imposing a transverse-momentum threshold on the target tracks. This includes the dense and challenging low-\(p_T\) regime, which constitutes the largest fraction of the charged-particle content of the event and substantially increases the combinatorial complexity of the reconstruction problem. More than 90\% of the considered detector hits originate from particles with \(p_T<1\,\mathrm{GeV}\). Moreover, the TrackML simulation propagates charged particles only above \(p_T=150\,\mathrm{MeV}\); the physical charged-particle population extends to substantially lower momenta, although such particles may not necessarily produce sufficiently extended trajectories for conventional track reconstruction.
Combining sparse full-event attention with learned local refinement, AlphaHits-TrackFormer achieves an inclusive TrackML score of 0.959 across this full population.

Building on the tracking architecture, AlphaHits extends
hit-level learning to multi-task event reconstruction and
classification through tracking-based pretraining and
joint fine-tuning. From shared detector-level representations,
it predicts tracks, jets, jet flavour, and event-level
observables. For Higgs event classification, AlphaHits achieves an AUC of 0.932 using tracker hits alone and 0.944 when including calorimeter towers, showing promising performance compared to a conventional tracker-plus-calorimeter reconstruction chain (AUC of 0.924).

Together, these results establish scalable full-event
tracking and multi-task learning as complementary steps
towards general-purpose hit-level foundation  models and Large Physics Models.
With larger-scale training and further developments, such approaches could
ultimately enable end-to-end optimization of physics
analyses directly from detector measurements, potentially
improving sensitivity beyond conventional reconstruction
pipelines.

\section*{LLM assistance disclosure}

This study builds on our previous implementations of TrackFormer and HiggsFormer. The new implementations were developed with the assistance of large language models, including GPT-5, GPT-6, and Claude Opus, primarily for code generation, debugging, and implementation support. The LLM-assisted code and intermediate results were reviewed and checked by the authors. Some figures included in this publication were generated with assistance from Claude. The scientific interpretation and conclusions of the manuscript were written and reviewed by the authors, with parts of the text subsequently improved using LLM-based editing. Parts of the Appendix were initially distilled from the implementation code using LLMs and were subsequently reviewed, corrected, and edited by the authors.

\section*{Acknowledgements} 
The author(s) gratefully acknowledge the computer resources at Artemisa, funded by the European Union ERDF and Comunitat Valenciana as well as the technical support provided by the Instituto de Física Corpuscular, IFIC (CSIC-UV). The work of R. RdA was supported by PID2023-148162NB-C22 from the Spanish Ministerio de Ciencia e Innovación and by the PROMETEO/2022/69 from the Spanish GVA. 
The work of J. Sohrabi was supported by the Generalitat Valenciana under the Santiago Grisolía programme (CIGRIS/2023/083). The author also acknowledges José David Martín Guerrero.

\appendix

\section{AlphaHits-TrackFormer}
\subsection{AlphaHits-TrackFormer Hyperparameters}
All training of the AlphaHits-TrackFormer models was conducted on an NVIDIA H200 card, and the training for the best model took roughly 1 week.

\subsubsection{Global stage}

The Router is first pretrained for one epoch using only the supervised routing objective. The Router and Tracker are subsequently trained jointly for 50 epochs using the AdamW optimizer with an initial learning rate of $3\times10^{-4}$, weight decay of $1\times10^{-4}$, and cosine learning-rate annealing. The batch size is one complete event, including all detector hits.

Training follows a staged routing curriculum that gradually transitions from target-guided to fully learned routing. During the first three epochs of joint training, the target routing assignment is enforced as the primary route. During the following three epochs, target-guided routing is used with a probability of 0.5. For all subsequent epochs, the routing assignments are determined entirely by the learned Router. Each valid detector hit is represented in two routing groups throughout training.

The global stage is trained with a multi-task objective combining discrete track-parameter prediction, hit classification, continuous track-parameter regression, charge prediction, and routing supervision:
\[
\begin{aligned}
L_{\mathrm{total}} ={}&
1.0 L_{\mathrm{token}}
+0.5 L_{\mathrm{signal}}
+1.0 L_{\mathrm{reg}}\\
&+0.2 L_{\mathrm{charge}}
+0.5 L_{\mathrm{router}}
+0.01 L_{\mathrm{balance}}.
\end{aligned}
\]

The token loss,
\[
L_{\mathrm{token}}
=
\frac{1}{3}
\left(
L_{p_T}
+
L_{\eta}
+
L_{\phi}
\right),
\]
supervises the three discrete track-parameter predictions. Instead of standard hard-label cross-entropy, we use neighbor-smoothed ordinal cross-entropy, in which a small fraction of the target probability is assigned to the immediately adjacent bins. The smoothing coefficients are 0.04, 0.06, and 0.08 for $p_T$, $\eta$, and $\phi$, respectively. For $\phi$, the first and last bins are treated as neighbors because the azimuthal coordinate is periodic and the directions at $-\pi$ and $+\pi$ are identical. The $p_T$ loss additionally uses class weighting to reduce the dominance of the highly populated low-\(p_T\) bins in the strongly non-uniform track-momentum spectrum.

The signal term $L_{\mathrm{signal}}$ is a cross-entropy loss for the binary classification of hits as belonging to a reconstructable particle or to background. Similarly, $L_{\mathrm{charge}}$ is a cross-entropy loss for predicting the sign of the particle charge for signal hits.

The continuous regression loss supervises the eight predicted quantities
\[
(\log p_T,\eta,\sin\phi,\cos\phi,
v_z,d_0,z_{\mathrm{POCA}},q/p_T).
\]
Huber losses are used throughout. The first four quantities form the principal regression objective, while the remaining four quantities provide auxiliary track-parameter supervision:
\[
\begin{aligned}
L_{\mathrm{reg}} ={}&
L_{\mathrm{Huber}}
(\log p_T,\eta,\sin\phi,\cos\phi)\\
&+0.20L_{v_z}
+0.15L_{d_0}
+0.20L_{z_{\mathrm{POCA}}}
+0.15L_{q/p_T}.
\end{aligned}
\]
The auxiliary quantities improve the physical information contained in the per-hit representation but do not enter the discrete candidate-track identifier.

The Router is supervised through $L_{\mathrm{router}}$, a soft-target cross-entropy loss over the 512 routing groups. For hits associated with reconstructable truth particles, the routing targets are constructed from $(r_{\mathrm{hit}},\eta_{\mathrm{track}},\phi_{\mathrm{track}})$, whereas non-signal hits use their hit-level $(r_{\mathrm{hit}},\eta_{\mathrm{hit}},\phi_{\mathrm{hit}})$ coordinates. The additional term $L_{\mathrm{balance}}$ regularises the mean routing probabilities and discourages the Router from concentrating an excessive fraction of hits in a small number of routing groups.
Following the importance-balancing regularisation introduced for sparsely gated
mixture-of-experts models~\cite{shazeer2017outrageously}, this term is added:
\[
L_{\mathrm{balance}}
=
G\sum_{g=1}^{G}
\left(
\bar{p}_g-\frac{1}{G}
\right)^2,
\]
where \(G=512\) is the number of routing groups and
\(\bar{p}_g\) denotes the routing probability for group \(g\), averaged over all
valid hits in the batch. The loss is minimised when the average routing
probability is uniformly distributed across the routing groups.

The Tracker consists of 18 Transformer encoder blocks with eight attention heads each, an embedding dimension of 256, a feedforward dimension of 1536, and a dropout rate of 0.1.

\subsubsection{Local stage}

The local stage consists of deterministic charge-aware preconditioning, a learned Gate, and the local slot-based Refiner.

The charge-aware preconditioning is deterministic and operates on the candidate groups produced by the global Tracker. A candidate is split by predicted charge when at least 50\% of its retained hits have a charge confidence above 0.8 and at least two confident hits are present for each charge sign. Small fragments containing at most two hits may subsequently be merged into a charge-compatible directly neighboring candidate in $(p_T,\eta,\phi)$ space. A merge is accepted only if the target contains at least four hits, is at least twice as large as the source fragment, and is sufficiently preferred over the second-best neighboring candidate. These conservative requirements improve the initial candidate assignment while limiting erroneous deterministic merges.

The Gate is trained to predict whether evaluating the Refiner around a candidate centre will contribute to the final reconstruction. Training labels are obtained by running the full Refiner: a centre is labelled as productive if a proposal originating from that centre survives proposal selection and changes the final hit assignment. The Gate uses a weighted binary cross-entropy-with-logits loss to account for the imbalance between productive and non-productive centres. It is trained for 12 epochs using AdamW with a learning rate of $1\times10^{-3}$, weight decay of $1\times10^{-4}$, and a batch size of 8192 candidate centres.

The Refiner is trained for 12 epochs using AdamW with an initial learning rate of $1.5\times10^{-4}$, weight decay of $1\times10^{-4}$, and cosine learning-rate annealing. Each training batch contains eight local neighborhoods.

Since the Refiner slots have no predefined identity, predicted slots must first be associated with truth particles. For each neighborhood, Hungarian matching is used to find a one-to-one assignment between truth tracks and predicted slots. 
The matching cost combines the hit-assignment likelihood and the predicted track parameters in order to associate each truth track with the most compatible Refiner slot. Because the Refiner slots are permutation-invariant and have no predefined identity, this correspondence is determined using Hungarian bipartite matching~\cite{kuhn1955hungarian,carion2020end}.

For a predicted slot \(k\) and a truth track \(t\), we define the matching cost as
\[
C_{kt}
=
C^{\mathrm{assign}}_{kt}
+
0.2\,C^{\mathrm{param}}_{kt}.
\]
The assignment component \(C^{\mathrm{assign}}_{kt}\) is the TrackML-weighted cross-entropy evaluated over the hits belonging to truth track \(t\), measuring how strongly these hits are assigned to predicted slot \(k\).

The parameter term \(C^{\mathrm{param}}_{kt}\) is the mean Smooth-L1 (Huber) loss between the predicted and true track parameters.
 The Hungarian algorithm then selects the minimum-cost one-to-one assignment between truth tracks and Refiner slots.

After the slot-to-truth correspondence has been established, the Refiner is optimized with
\[
L_{\mathrm{Refiner}}
=
L_{\mathrm{assign}}
+0.25L_{\mathrm{exist}}
+0.20L_{\mathrm{param}}
+0.30L_{\mathrm{track}}.
\]
Here, $L_{\mathrm{assign}}$ is a TrackML-weighted cross-entropy loss for assigning each hit either to its matched track slot or to the background class. $L_{\mathrm{exist}}$ is a binary cross-entropy loss indicating which of the 32 slots correspond to a truth track, and $L_{\mathrm{param}}$ is a Smooth-L1 loss on the continuous parameters of matched slots.

The final term, ($L_{\mathrm{track}}$), is a soft track-level purity-and-coverage objective based on a differentiable Dice/F1 overlap~\cite{sudre2017generalised}. It encourages matched slots to recover the hits of a truth track while limiting contamination from other particles.
Its individual contribution to the reconstruction performance has not been studied by us.

\subsection{Binning strategies}

\label{sec:binning}

The Tracker gives every hit a discrete token, and hits with the same token form one
candidate track. We build this token by binning $\pt$, $\eta$ and $\phi$ separately and
combining the three bin indices. Three heads with about 200 outputs per hit then address
about $1.6\times10^5$ tokens, because the cost grows with the sum of the bin counts and
the number of tokens with their product~\cite{Jegou2011PQ,Mentzer2024FSQ}, and
neighbouring bins describe neighbouring trajectories. The obvious alternative is a token
space learned by a vector-quantised variational autoencoder
(VQ-VAE)~\cite{vandenOord2017VQVAE}, which needs one output per code and numbers its codes
arbitrarily.

\begin{figure}[H]
\centering
\begin{minipage}[c]{0.55\linewidth}
\centering
\small
\captionof{table}{Factorised binning and a VQ-VAE codebook in the same 14-layer
Tracker, on 874 validation events with 200--500 particles each.}
\label{tab:binning}
\begin{tabular}{@{}lcc@{}}
\toprule
 & Binning & VQ-VAE \\
\midrule
Number of tokens & $\approx 1.6\times10^5$ & 12\,000 \\
Outputs per hit & $\approx 200$ & 12\,000 \\
Tracker parameters & 15.2\,M & 18.2\,M \\
TrackML, Tracker & 0.926 & 0.933 \\
TrackML, Tracker + Refiner & \textbf{0.989} & -- \\
Charge accuracy & 98.8\% & 98.6\% \\
Vertex $z$ mean abs.\ error & 22.0\,mm & 26.1\,mm \\
\bottomrule
\end{tabular}

\vspace{2pt}
{\footnotesize TrackML scores for $\pt>0.9$\,GeV.}
\end{minipage}\hfill
\begin{minipage}[c]{0.42\linewidth}
\centering
\includegraphics[width=\linewidth]{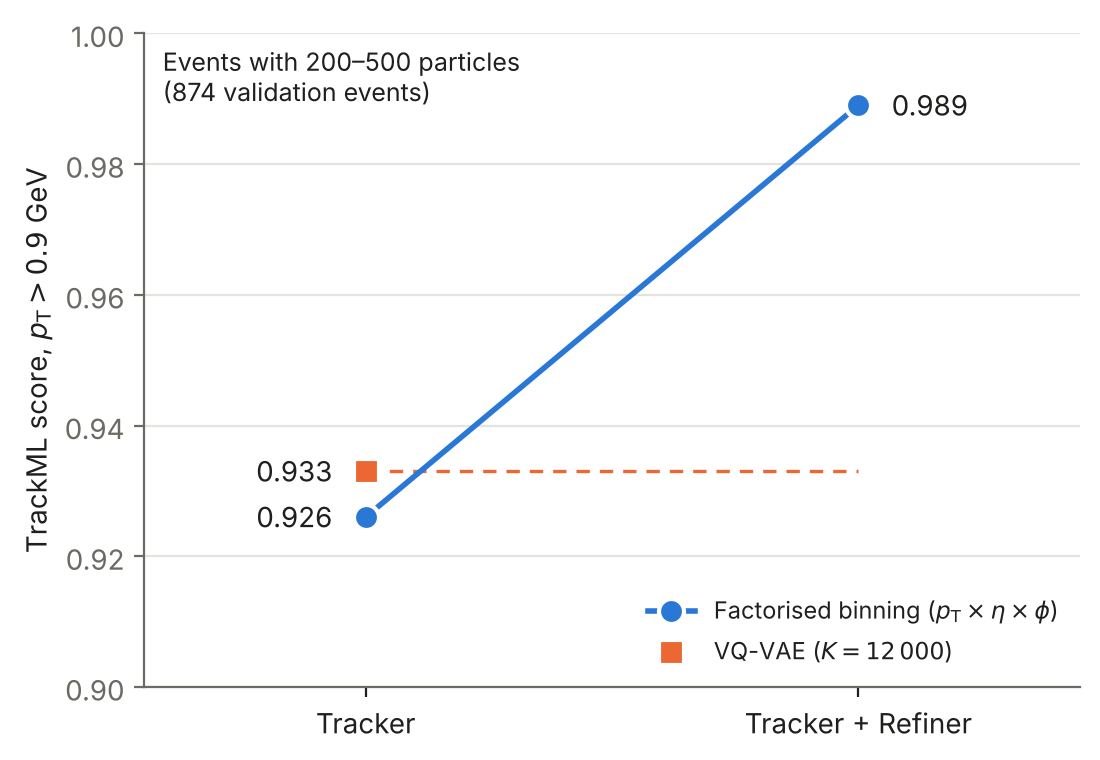}
\captionof{figure}{TrackML score for $\pt>0.9$\,GeV before and after local refinement,
on 874 validation events with 200--500 particles each. The Refiner needs ordered bins
and applies to the binning only.}
\label{fig:binning-vqvae}
\end{minipage}
\end{figure}

We trained a 14-layer Tracker twice on events with 200--500 particles (7\,871
training and 874 validation events): once with the binning, and once with a VQ-VAE
codebook of $K=12\,000$ codes fitted to the charge-free kinematics of the training
particles. At the Tracker stage the two are close (Table~\ref{tab:binning}). The VQ-VAE
scores 0.933 against 0.926 for $\pt>0.9$\,GeV,
using 3\,M more parameters in its output head. The two separate after the Tracker. When
the factorised Tracker gets a bin wrong, it is almost always off by one (96--99\% of hits
are within one bin on each axis), so the hits of a split particle land in neighbouring
bins. The Refiner takes the hits in a block of $3\times3\times5$ bins around each
candidate and reassigns them to tracks, which raises the score to 0.989
(Figure~\ref{fig:binning-vqvae}). VQ-VAE codes have no neighbours in this sense, so the
learned codebook stays at 0.933.

The test events are also sparse, with about 350 particles each against
$\mathcal{O}(10^4)$ in a full TrackML event, and particles with the same token merge into
one candidate. If $N$ particles are spread uniformly over $K$ tokens, a fraction of about
$1-e^{-N/K}$~\cite{Mitzenmacher2017Probability} share their token with another particle:
3\% for the VQ-VAE and 0.2\% for the binning in the test events, but 57\% and 6\% in a
full event, where the binning value is a lower bound because real bins are filled
unevenly. Matching the binning would take about $1.6\times10^5$ codes and an output head
of 41\,M parameters, 2.7 times the rest of the Tracker, and codebooks of this size tend
to leave many codes unused~\cite{Lancucki2020RobustVQ,Zhu2024VQGANLC}. These are
estimates, since we have not compared the two tokenizers on full events, but they agree
with the measurements above, and AlphaHits-TrackFormer therefore uses factorised binning.

\subsection{AlphaHits-TrackFormer scaling with model depth}
\begin{figure}[H]
    \centering
    \includegraphics[scale=0.6]{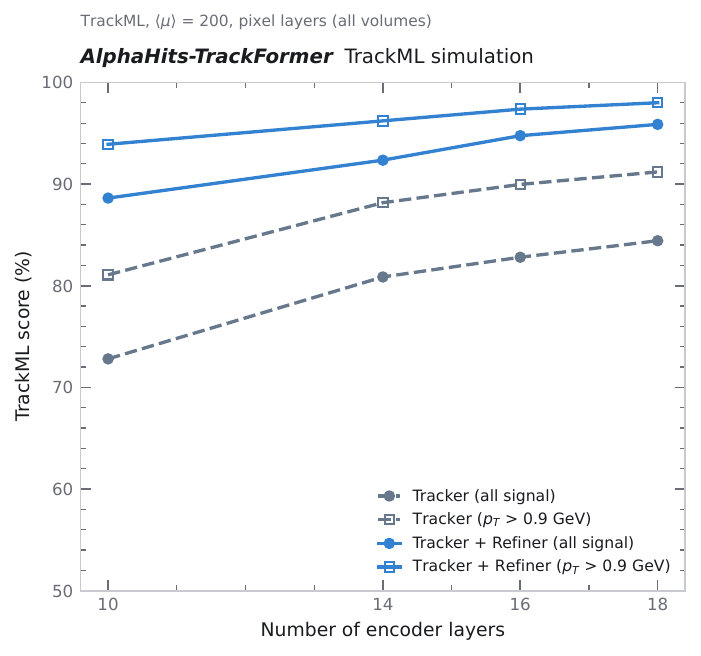}
    \caption{TrackML score of the full AlphaHits-TrackFormer pipeline scaling with model size of the Tracker. Results are shown both for the Tracker alone and after the preconditioning, gating and refinements are applied. The largest and best-performing model, the 18-layer AlphaHits-TrackFormer has just over 19 million parameters.}
    \label{fig:tracking-scaling}
\end{figure}

\subsection{AlphaHits-TrackFormer differential performance}
\begin{figure}[H]
    \centering
    \includegraphics[scale=0.45]{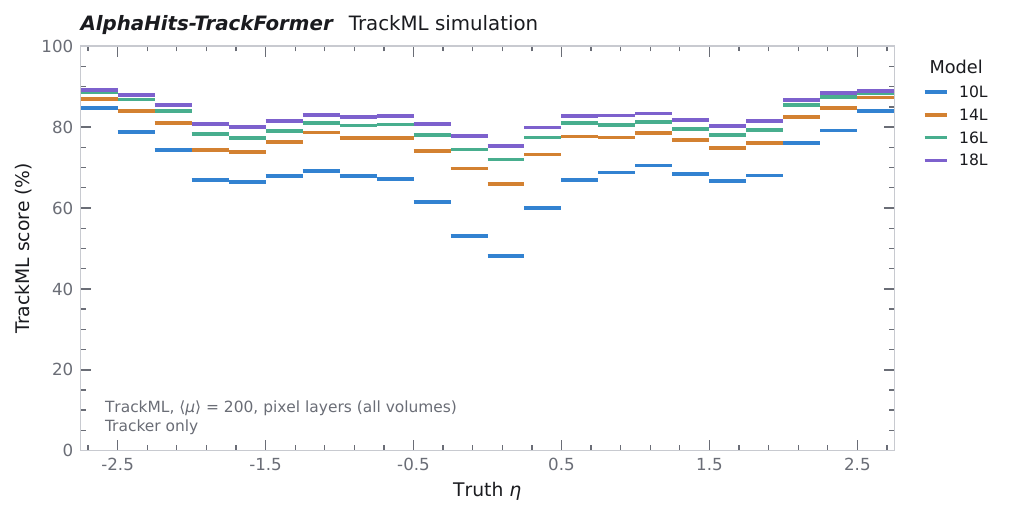}
    \includegraphics[scale=0.45]{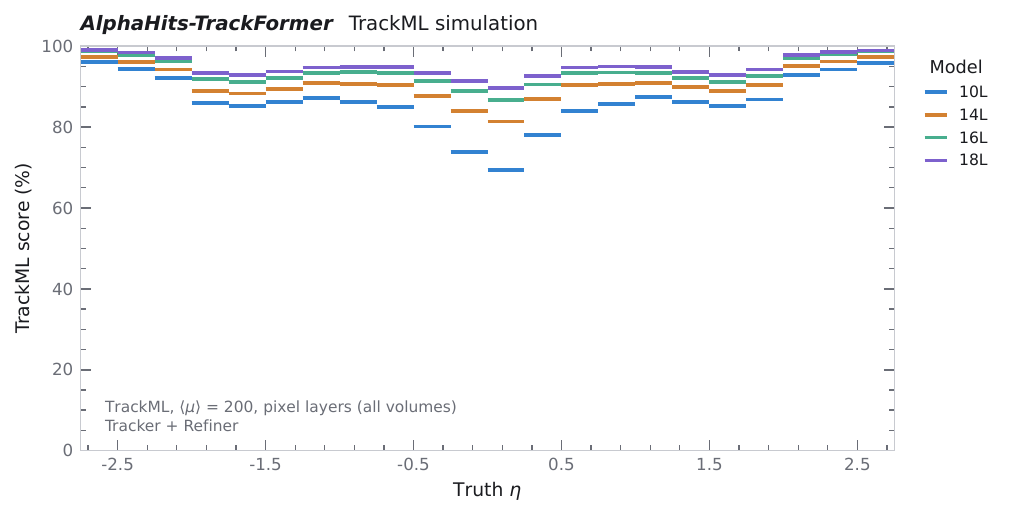}
    \includegraphics[scale=0.45]{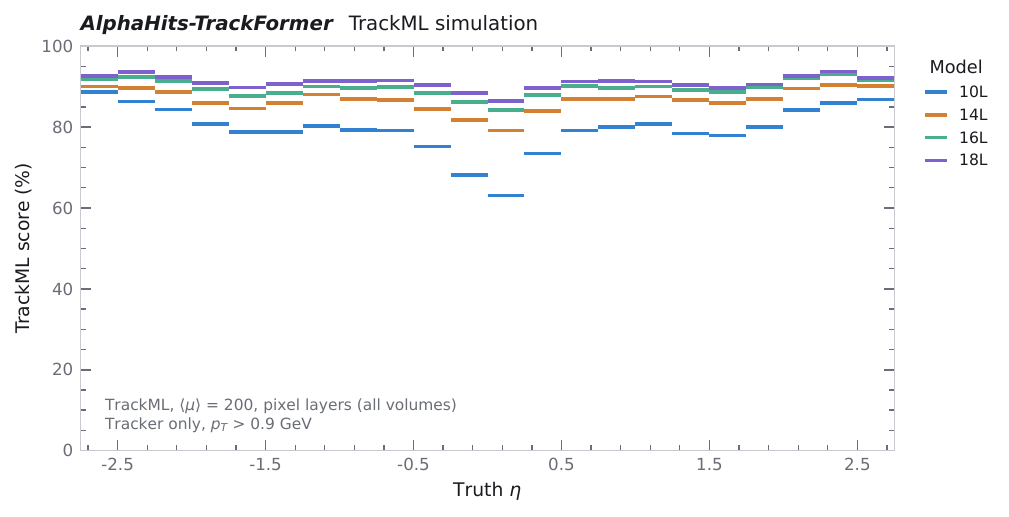}
    \includegraphics[scale=0.45]{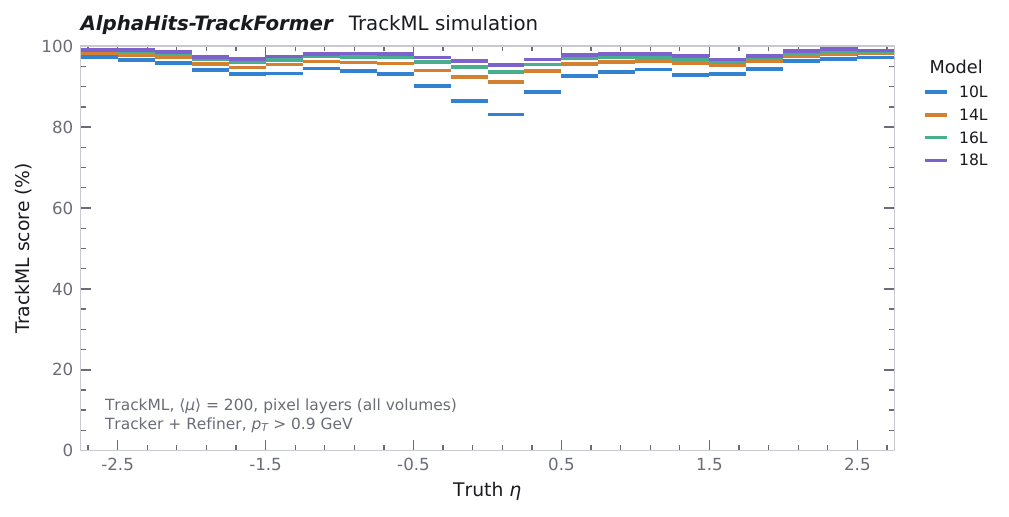}
    \caption{AlphaHits-TrackFormer TrackML score as a function of particle $\eta$ for Tracker (left) and Tracker+Refiner (right).}
    \label{fig:differential_eta}
\end{figure}

\begin{figure}[H]
    \centering
    \includegraphics[scale=0.45]{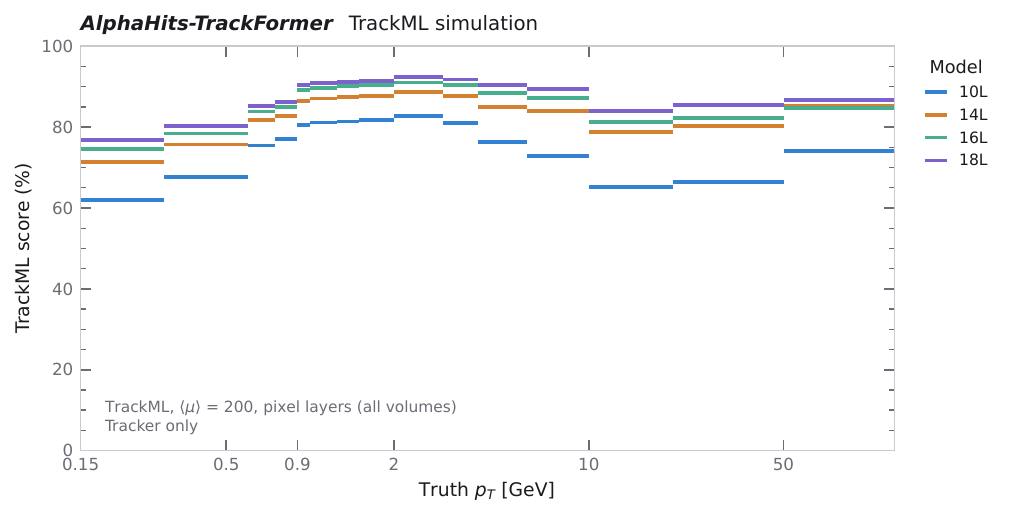}
    \includegraphics[scale=0.45]{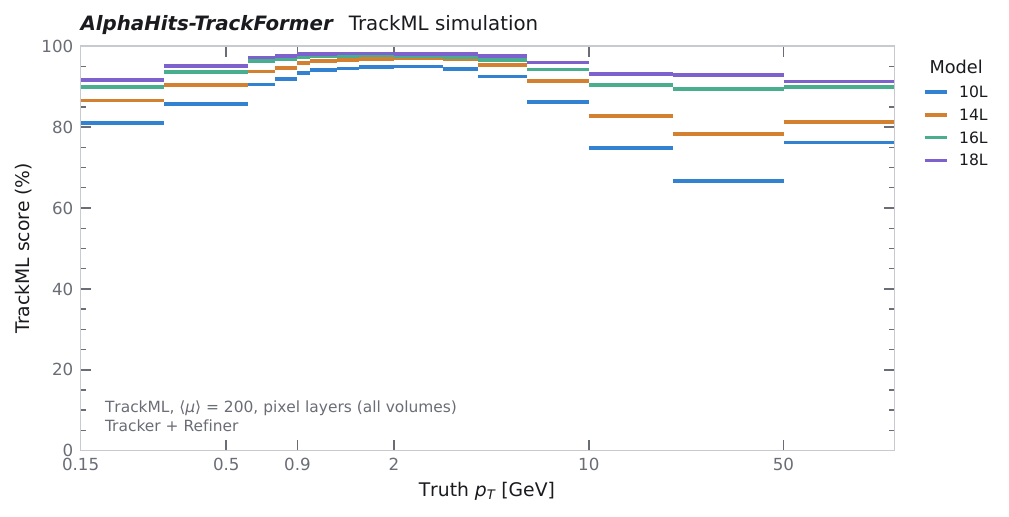}
    \caption{AlphaHits-TrackFormer TrackML score as a function of particle $\pt$ for Tracker (left) and Tracker+Refiner (right).}
    \label{fig:differential_pt}
\end{figure}

\subsection{AlphaHits-Trackformer reconstruction ladder performance}
\begin{table}[H]
  \centering\small
  \setlength{\tabcolsep}{6pt}\renewcommand{\arraystretch}{1.15}
  \caption{AlphaHits-TrackFormer reconstruction ladder for the 18-layer Tracker, with the full scan of the Gate threshold. Upper block: TrackML score on the 820 test events, over all particles, for $p_{\mathrm{T}}>0.9$\,GeV and in slices of $|\eta|$ (for $p_{\mathrm{T}}>0.9$\,GeV). The fraction of candidate centres sent to the Refiner and the mean time per event (Router and Tracker plus the local stage) are measured on 50 of the events on an NVIDIA H100, with an inference batch of 256 neighbourhoods.}
  \label{tab:tf:ladder_full}
  \begin{tabular}{@{}lcccccccc@{}}
    \toprule
    Stage & all & $p_{\mathrm{T}}>0.9$ & $|\eta|<1$ & $1\le|\eta|<2$ & $2\le|\eta|<3$ & $|\eta|\ge3$ & centres refined & time [s] \\
    \midrule
    Tracker & 0.844 & 0.912 & -- & -- & -- & -- & 0\,\% & 0.45 \\
    + preconditioner & 0.880 & 0.934 & 0.923 & 0.927 & 0.955 & 0.935 & 0\,\% & 0.47 \\
    full pipeline, no Gate & 0.959 & 0.981 & 0.973 & 0.975 & 0.992 & 0.989 & 100\,\% & 4.60 \\
    Gate 0.10 & 0.959 & 0.981 & 0.973 & 0.975 & 0.992 & 0.989 & 69\,\% & 3.48 \\
    Gate 0.20 & 0.959 & 0.981 & 0.973 & 0.975 & 0.991 & 0.989 & 61\,\% & 3.17 \\
    Gate 0.30 & 0.959 & 0.980 & 0.972 & 0.974 & 0.991 & 0.989 & 56\,\% & 2.98 \\
    Gate 0.50 & 0.957 & 0.977 & 0.969 & 0.971 & 0.989 & 0.987 & 51\,\% & 2.75 \\
    \bottomrule
  \end{tabular}
\end{table}

\subsection{AlphaHits-Trackformer latency}
\begin{figure}[H]
    \centering
    \includegraphics[scale=0.8]{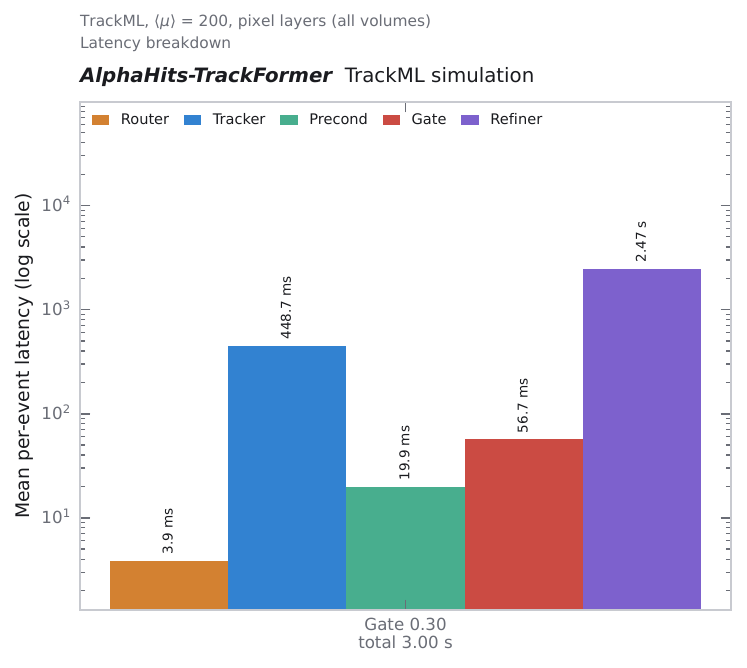}
    \caption{The average latency at inference time of the 18-layer AlphaHits-TrackFormer pipeline, broken down per stage. The Gate model is used with threshold of 0.30.}
    \label{fig:latency}
\end{figure}

The inference time of the AlphaHits-TrackFormer was evaluated on an NVIDIA H100 GPU card. The latency broken down per stage is reported in Figure~\ref{fig:latency}, when using the Gate model with threshold of 0.30. In total, the pipeline takes 3s to process a single event. The target of this effort was accomplishing a pipeline that can process the full TrackML high-luminosity events and solve the tracking challenge well. Therefore, fast inference was not yet among the primary goals when designing the AlphaHits-TrackFormer.

\section{AlphaHits}
\subsection{AlphaHits Data Generation}
\begin{table}[H]
  \centering\small
  \setlength{\tabcolsep}{9pt}\renewcommand{\arraystretch}{1.15}
  \caption{The AlphaHits Dataset Samples.}
  \label{tab:ah:sample}
  \begin{tabular}{@{}ll@{}}
    \toprule
    Processes & $t\bar{t}H$ ($H\to b\bar b$) signal and $t\bar{t}$ background, 100\,000 events each; no pileup \\
    Generator & \textsc{Pythia} 8.311 in ACTS 41.1, $\sqrt{s}=14$\,TeV, inclusive top and $W$ decays \\
    Parton-level cut & $\hat p_{\mathrm{T}}>100$\,GeV on the hard process \\
    Detector & ACTS Fatras, silicon-only GenericDetector (TrackML layout), 2\,T \\
    Hits & digitised cluster centroids; 2\,615 ($t\bar{t}H$), 2\,290 ($t\bar{t}$) per event; at most 6\,245 \\
    GenJets ($p_{\mathrm{T}}>20$\,GeV) & 6.8 per event \\
    Trackable particles ($\ge3$ hits) & 193 per event, carrying 90\,\% of the hits \\
    Towers (calorimeter variant) & 261 per event on average, at most 614 \\
    References and targets & \textsc{Delphes} 3.5.1, ATLAS-like card; full chain and tracker-only chain \\
    Split per process & 80\,000 / 10\,000 / 10\,000 training / validation / test \\
    \bottomrule
  \end{tabular}
\end{table}

\subsection{Delphes $b$-tagging reference}\label{app:ah:btag}
Delphes tags each jet with a probability that depends only on its true flavour and $\pt$ (GeV): $\varepsilon_b = A\,\tanh(0.003\,\pt)\,30/(1+0.086\,\pt)$, $\varepsilon_c = C\,\tanh(0.02\,\pt)/(1+0.0034\,\pt)$ and $\varepsilon_{\mathrm{light}} = L_0 + L_1\,\pt$, the ATLAS-card parametrisation of Delphes~3.5.1~\cite{Delphes2014}. The card's five presets, named 85, 77, 70, 60 and 40, tag fewer $b$ jets than their names say (the 70\,\% preset, $A=0.80$, tags 66\,\% of the $t\bar{t}$ $b$ jets). For Figure~\ref{fig:ah:ftag} the amplitude $A$ was retuned to measured efficiencies of 40, 60, 70, 77, 80 and 85\,\% and the events were re-simulated; $C$, $L_0$ and $L_1$ follow the card's own dependence on $A$ between the presets (Table~\ref{tab:ah:btag}), an extrapolation above the last preset. Each tag is an independent draw per jet and working point. No track quality or vertexing enters: the reference measures the card's parametrisation, not a tagger.

\begin{table}[H]
  \centering\small
  \caption{Delphes $b$-tagging settings of Figure~\ref{fig:ah:ftag}. Measured $b$-efficiency on the 20k test events, graded against the truth flavour, for calorimeter jets / track jets.}
  \label{tab:ah:btag}
  \begin{tabular}{@{}lcccc cc@{}}
    \toprule
    working point & $A$ & $C$ & $L_0$ & $L_1$ [GeV$^{-1}$] & $\varepsilon_b$ calo & $\varepsilon_b$ tracks \\
    \midrule
    40\,\% & 0.485 & 0.022 & $3.8\times10^{-4}$ & $1.3\times10^{-6}$ & 0.418 & 0.407 \\
    60\,\% & 0.727 & 0.070 & $1.4\times10^{-3}$ & $4.6\times10^{-6}$ & 0.627 & 0.606 \\
    70\,\% & 0.849 & 0.177 & $7.7\times10^{-3}$ & $1.8\times10^{-5}$ & 0.729 & 0.709 \\
    77\,\% & 0.934 & 0.331 & $2.4\times10^{-2}$ & $4.0\times10^{-5}$ & 0.806 & 0.779 \\
    80\,\% & 0.970 & 0.430 & $3.8\times10^{-2}$ & $5.6\times10^{-5}$ & 0.837 & 0.814 \\
    85\,\% & 1.031 & 0.672 & $8.5\times10^{-2}$ & $1.0\times10^{-4}$ & 0.889 & 0.857 \\
    \bottomrule
  \end{tabular}
\end{table}

\subsection{Summary of training stages of AlphaHits model}
\begin{table}[H]
  \centering\small
  \setlength{\tabcolsep}{9pt}\renewcommand{\arraystretch}{1.15}
  \caption{The two training stages of AlphaHits. Both use the 160\,000-event training split, AdamW (weight decay $10^{-4}$), batch 64 as 4 GPUs $\times$ 16 events under distributed data parallelism, float16 mixed precision, seed 42, a reduce-on-plateau schedule ($\times0.5$, patience 20) and early stopping (patience 60, minimum improvement $10^{-4}$) on the validation loss; the checkpoint is selected on validation. Stage 2 copies 407 of 496 tensors (9.74\,M parameters) from the stage-1 checkpoint; the 89 head tensors (2.71\,M) start fresh. The towers variant is stage 2 only, with the same recipe and stage-1 checkpoint plus the gated cross-attention adapters; its early stopping fired at epoch 466.}
  \label{tab:ah:training}
  \setlength{\tabcolsep}{7pt}
  \begin{tabular}{@{}lll@{}}
    \toprule
     & Stage 1: tracker pre-training & Stage 2: joint fine-tuning \\
    \midrule
    Heads trained & tracking, hit particle & all nine \\
    Loss & $L_{\mathrm{trk}}+0.4\,L_{\mathrm{hitpart}}$ & geometric mean of six tasks + auxiliary terms \\
    Initialisation & random & stage-1 checkpoint, fresh optimiser \\
    Trainable parameters & 9.74\,M & 12.45\,M \\
    Initial learning rate & $3\times10^{-5}$ & $1.25\times10^{-5}$ \\
    Plateau reductions & 2 (epochs 375, 456) & 5 (epochs 156 to 462) \\
    Final learning rate & $7.5\times10^{-6}$ & $3.9\times10^{-7}$ \\
    Early-stopping monitor & validation loss & median of the last 5 validation losses \\
    Epochs run & 500 of 500 & 500 of 500 \\
    Selected checkpoint & epoch 500 & epoch 489 \\
    Wall time, $4\times$ H100 & 41\,h (4.9\,min per epoch) & 58\,h (6.9\,min per epoch) \\
    \bottomrule
  \end{tabular}
\end{table}

\subsection{Targets and loss}
\label{sec:hf:targets}
AlphaHits is supervised on nine targets (Table~\ref{tab:ah:heads}), derived
from the generator record or the Delphes full chain of the same event. Regression targets are standardised with training-split moments,
\begin{equation}
  \label{eq:hf:norm}
  z=\frac{\log(1+x)-\mu}{\sigma}\ \ (\text{positive, right-skewed }x),\qquad
  z=\frac{x-\mu}{\sigma}\ \ (\eta\text{-like }x),\qquad
  \phi\mapsto(\sin\phi,\cos\phi),
\end{equation}
so that one loss scale serves every regression; per-hit targets carry a validity
mask $m\in\{0,1\}$ and their moments are fitted on valid entries only.

Per-task losses $L_t$ are combined as a geometric mean over the \emph{main}
tasks plus fixed-weight \emph{auxiliary} terms,
\begin{align}
  \label{eq:hf:total1}
  \mathcal L^{(1)} &= L_{\mathrm{trk}} + 0.4\,L_{\mathrm{hitpart}},\\
  \label{eq:hf:total2}
  \mathcal L^{(2)} &= \big(L_{\mathrm{cls}}\,L_{\mathrm{HT}}\,L_{\mathrm{MET}}\,L_{\mathrm{lep}}\,L_{\mathrm{jets}}\,L_{\mathrm{trk}}\big)^{1/6}
     + 0.2\,L_{\mathrm{nhits}} + 0.4\,L_{\mathrm{hitpart}} + 0.3\,L_{\mathrm{hitjet}},
\end{align}
for stage~1 (Tracker backbone) and stage~2 (one-shot reconstruction). Both instantiate $\mathcal L=\big(\prod_{t\in\mathcal M}L_t\big)^{1/|\mathcal M|} +\sum_{a\in\mathcal A}w_aL_a$, evaluated as $\exp\big[\frac{1}{|\mathcal M|}\sum_t\log\max(L_t,10^{-6})\big]$. With the single main task of stage~1 the mean reduces to $L_{\mathrm{trk}}$ itself. The geometric mean is scale-free, $\partial\mathcal L/\partial L_t=\mathcal L/(|\mathcal M|\,L_t)$: main tasks need no weights, and a task whose loss is already small receives the larger relative gradient. Auxiliary terms stay outside the  product, so a scaffold loss near zero cannot collapse it. Table~\ref{tab:hf:losses} summarises all terms.

\begin{table}[H]
  \centering
  \caption{Loss terms of AlphaHits, their weight or role in the combination of
    Eqs.~\eqref{eq:hf:total1}--\eqref{eq:hf:total2}, and the training stage in
    which they are active. ``Main'' terms enter the geometric mean without a
    weight; ``aux'' terms are added with the quoted weight. Within the jet loss
    the three parts are combined with weights $(1,5,1)$ and a no-object weight
    $w_{\varnothing}=0.85$ on unmatched queries (events without truth jets use
    the unweighted existence BCE). All Huber terms use $\delta=1$ on
    standardised targets; all cross-entropies are means over the valid entries
    of their mask.}
  \label{tab:hf:losses}
  {\footnotesize\setlength{\tabcolsep}{4pt}\renewcommand{\arraystretch}{1.1}
  \begin{tabular}{@{}p{0.22\linewidth}p{0.20\linewidth}p{0.17\linewidth}p{0.26\linewidth}c@{}}
  \toprule
  Quantity & Loss & Weight / role & Responsible for & Stage \\
  \midrule
  event class & BCE & main & signal vs.\ background & 2 \\
  $H_{\mathrm{T}}$, $E_{\mathrm{T}}^{\mathrm{miss}}$ & Huber & main & hadronic and invisible energy scales & 2 \\
  $n_\ell$, leading $p_{\mathrm{T}}^{\ell}$ & Huber $+$ Huber & main & leptonic decay content & 2 \\
  jet existence & weighted BCE, $w_{\varnothing}=0.85$ & main (jets), $\times1$ & jet multiplicity, operating point & 2 \\
  jet $(p_{\mathrm{T}},\eta,\phi)$ & $L_1$ on matched pairs & main (jets), $\times5$ & jet finding and energy & 2 \\
  jet flavour & 4-way CE on matched pairs & main (jets), $\times1$ & $b$-tagging & 2 \\
  track tokens $(p_{\mathrm{T}},\eta,\phi)$ & $3\times$ masked CE, averaged & main (tracking) & per-hit track-parameter field & 1, 2 \\
  trackable bit & BCE over real hits & main (tracking), $+1$ & which hits belong to a track & 1, 2 \\
  particle $(p_{\mathrm{T}},\eta,\phi,\rho,v_z)$ & masked Huber & aux, $+0.4$ & continuous track parameters, displacement & 1, 2 \\
  particle charge & 3-way masked CE & aux (inside hitpart) & bending direction & 1, 2 \\
  member jet $(\Delta\eta,\Delta\phi,$ $p_{\mathrm{T}},m,L_{xy})$ & masked Huber & aux, $+0.3$ & hit-to-jet association, SV reach & 2 \\
  hit multiplicity & Huber & aux, $+0.2$ & occupancy scaffold & 2 \\
  \bottomrule
  \end{tabular}}
\end{table}

Classification uses the binary cross-entropy of the event logit $s$,
$\mathrm{BCE}(s,y)=-y\log\sigma(s)-(1-y)\log\big(1-\sigma(s)\big)$. Every
regression uses the Huber loss of the standardised residual $r=\hat z-z$,
\begin{equation}
  \label{eq:hf:huber}
  \ell_\delta(r)=\begin{cases}\tfrac12\,r^2, & |r|\le\delta,\\[2pt]
  \delta\,\big(|r|-\tfrac12\delta\big), & |r|>\delta,\end{cases}\qquad \delta=1,
\end{equation}
quadratic within about one standard deviation of the target and with the
gradient capped at $\pm\delta$ beyond it, so extreme events cannot dominate an
update. Hence $L_{\mathrm{HT}}$, $L_{\mathrm{MET}}$ and $L_{\mathrm{nhits}}$ are
batch means of $\ell_\delta$, and $L_{\mathrm{lep}}$ the sum of two ($n_\ell$ and
$p_{\mathrm{T}}^{\ell}$). Per-hit terms use the masked mean over hits $i$ and target
components $d$,
\begin{equation}
  \label{eq:hf:masked}
  \langle\ell\rangle_m=\frac{\sum_{i,d} m_{i,d}\,\ell_{i,d}}{\sum_{i,d} m_{i,d}},
\end{equation}
in which the mask enters numerator and denominator, so a padded,
particle-less, jet-less or non-trackable hit contributes exactly zero:
\begin{align}
  \label{eq:hf:hitpart}
  L_{\mathrm{hitpart}} &= \langle\ell_\delta\rangle_{m^{\mathrm{part}}}
                   + \langle\mathrm{CE}_{q}\rangle_{m^{\mathrm{part}}},\\
  \label{eq:hf:hitjet}
  L_{\mathrm{hitjet}}  &= \langle\ell_\delta\rangle_{m^{\mathrm{jet}}},\\
  \label{eq:hf:trk}
  L_{\mathrm{trk}}     &= \tfrac13\big(\langle\mathrm{CE}_{p_{\mathrm{T}}}\rangle_{m^{\mathrm{trk}}}
                   + \langle\mathrm{CE}_{\eta}\rangle_{m^{\mathrm{trk}}}
                   + \langle\mathrm{CE}_{\phi}\rangle_{m^{\mathrm{trk}}}\big)
                   + \langle\mathrm{BCE}_{\mathrm{trackable}}\rangle_{m^{\mathrm{hit}}},
\end{align}
where $m^{\mathrm{part}}$ marks hits with a particle record (one mask for all
six components), $m^{\mathrm{jet}}$ is component-wise ($L_{xy}$ additionally
requires a secondary vertex), $m^{\mathrm{trk}}$ marks trackable hits,
$m^{\mathrm{hit}}$ all real (non-padded) hits, $\mathrm{CE}_q$ is the
three-way charge cross-entropy and $\mathrm{CE}_{p_{\mathrm{T}},\eta,\phi}$ the
cross-entropies over the 32, 128 and 64 codebook bins.

Jets are a set prediction over $K=12$ queries, each emitting an existence
logit $s_k$, kinematics $\hat{\mathbf k}_k=(\hat p_{\mathrm{T}},\hat\eta,\widehat{\sin\phi},
\widehat{\cos\phi})$ and flavour probabilities $p_k(c)$. Truth jets
$(\mathbf k_i,c_i)$ are assigned to queries by a Hungarian matching of the cost
\begin{equation}
  \label{eq:hf:cost}
  C_{ki}=5\,\lVert\hat{\mathbf k}_k-\mathbf k_i\rVert_1-\log p_k(c_i)-\log\sigma(s_k),
\end{equation}
so kinematic distance, flavour likelihood and existence confidence all enter
the (non-differentiated) assignment. With $y_k=1$ for matched
queries, $w_k=1$ (matched) or $0.85$ (unmatched) and $\mathcal M$ the matched
pairs,
\begin{equation}
  \label{eq:hf:jets}
  L_{\mathrm{jets}}=\underbrace{\frac1K\sum_{k=1}^{K}w_k\,\mathrm{BCE}(s_k,y_k)}_{\text{existence}}
  \;+\;5\,\underbrace{\frac{1}{4|\mathcal M|}\sum_{(k,i)\in\mathcal M}\lVert\hat{\mathbf k}_k-\mathbf k_i\rVert_1}_{\text{kinematics}}
  \;+\;\underbrace{\frac{1}{|\mathcal M|}\sum_{(k,i)\in\mathcal M}-\log p_k(c_i)}_{\text{flavour}},
\end{equation}
averaged over events: the existence term decides how many jets exist (its
no-object weight sets the efficiency/fake operating point), the $L_1$ term
places and scales the matched ones, and the flavour term is the $b$-tagger,
$P(b)=p_k(b)$ being a jet's $b$-score at evaluation.

\subsection{Stage-1 tracking performance}
Table~\ref{tab:ah:tracking} scores the tracking head with the \TrackML{} metric on our events against the codebook ceiling of 0.9588. The stage-1 backbone alone reaches 0.891 with the plain equal-token decode and 0.927 with a validation-tuned decode, and a local Refiner of the AlphaHits-TrackFormer type trained on its output reaches the ceiling (0.959). After joint fine-tuning the same head gives 0.806 and 0.894, inside the forward pass that also returns the jets and the event class. Trained jointly from scratch, without stage~1, the recipe stays at 0.446 and its classifier at an AUC of 0.895: the pre-training is the recipe.

\begin{table}[!htb]
  \centering\small
  \setlength{\tabcolsep}{9pt}\renewcommand{\arraystretch}{1.15}
  \caption{Tracking performance of AlphaHits on the 20\,000 test events: TrackML score (weights $w_{p_{\mathrm{T}}}\,w_{\mathrm{order}}$, unit sum per event) for the stage-1 backbone alone, after the joint stage-2 fine-tuning, and for a model trained jointly from scratch without stage~1; with the plain decode (hits with equal tokens form a track) and a decode that merges neighbouring tokens and re-assigns hits with ambiguous token logits to a neighbouring track, its thresholds tuned on the validation split. The codebook ceiling, every hit given its truth token, is 0.9588; percentages are relative to it. The last column adds a local Refiner of the AlphaHits-TrackFormer type trained on the stage-1 output; it is an optional post-hoc stage, not part of the one-shot model.}
  \label{tab:ah:tracking}
  \begin{tabular}{@{}lccc@{}}
    \toprule
    Backbone & Equal-token decode & Tuned decode & + local Refiner \\
    \midrule
    Stage 1 alone (tracker pre-training) & 0.891 (93\,\%) & 0.927 (97\,\%) & 0.959 (100\,\%) \\
    Stage 2 (joint fine-tuning; the quoted model) & 0.806 (84\,\%) & 0.894 (93\,\%) & -- \\
    Joint training from scratch, no stage 1 & 0.446 (47\,\%) & 0.719 (75\,\%) & -- \\
    \bottomrule
  \end{tabular}
\end{table}

\subsection{Particle Transformer baseline training}
The reference classifier is the Particle Transformer (ParT)~\cite{Qu2022ParticleTransformer} with the standard backbone (8~layers, embedding dimension~128), trained from scratch for $t\bar{t}H$ versus $t\bar{t}$ on Delphes objects of the same generated events. Following our previous work~\cite{Builtjes2025PhysicsInteractions}, the model is used in its vanilla self-attention form with no pairwise features, and training uses the Ranger optimiser with learning rate $10^{-3}$ and batch size~512; the checkpoint with the best validation loss is kept.
Each event is a set of tokens with per-object features $(\log E,\log p_T,\eta,\phi)$ (with $\phi$ kept as the raw angle), a one-hot object identity, and an auxiliary $(\log E_T^{\mathrm{miss}},\phi_{\mathrm{miss}})$ input.
Two different object sets are used.
\emph{Delphes tracks}: anti-$k_t$ jets with $p_T>20$\,GeV, $R=0.6$ built from charged-particle tracks, each with $p_T$, $\eta$, $\phi$, $E$ and a binary $b$-tag, and the track-based $E_T^{\mathrm{miss}}$ (magnitude and direction), no leptons.
\emph{Delphes tracks + calorimeter}: anti-$k_t$ jets with $p_T>20$\,GeV, $R=0.6$ on calorimeter towers with the same features and $b$-tag, the calorimeter $E_T^{\mathrm{miss}}$, and electrons with $p_T$, $\eta$, $\phi$, $E$ and charge; muons are removed.
The $b$-tag is the Delphes parametrisation with its amplitude retuned to measured $b$-jet efficiencies of 80, 60 and 40\,\% in $t\bar{t}$.
The 10k, 20k and 40k points use 8\,000, 18\,000 and 38\,000 training events and 1\,000 validation events from the original 40\,000-event sample used for the Higgsformer~\cite{Caron2025Higgsformer} training and evaluation, generated with identical settings to the new 200\,000-event dataset.
The 160k point uses the 160\,000 training and 20\,000 validation events of the main sample.
Every ParT model is evaluated on the same 20\,000 test events as AlphaHits, which none of the training sets contain.
Each reported point is the mean of the three trainings.

\subsection{AlphaHits Performance Illustration}
\begin{figure}[H]
  \centering
  \includegraphics[width=\textwidth]{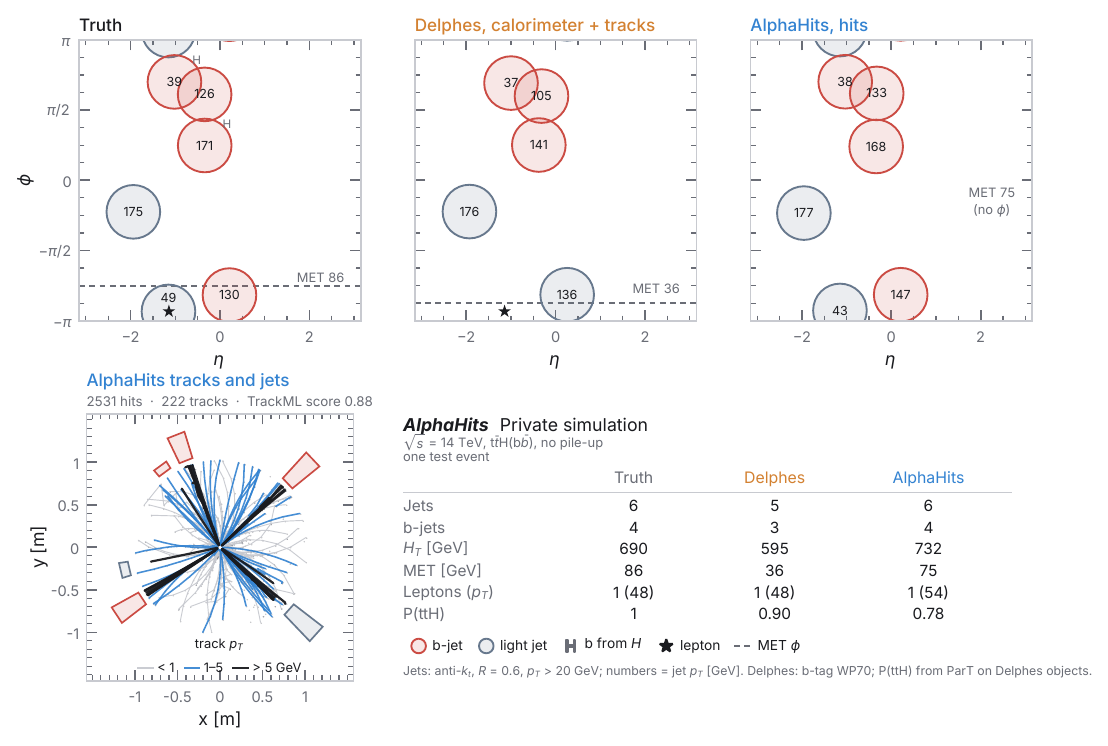}
  \caption{One $t\bar{t}H$ ($H\to b\bar b$) test event in $\eta$--$\phi$ as the truth record, as Delphes (calorimeter + tracks, $b$-tag WP70) and as AlphaHits from the hits (top), the hits with the AlphaHits reconstructed tracks and jets in the transverse plane (bottom left), and the event-level quantities: truth, Delphes and AlphaHits reconstruction (bottom right). Jets are anti-$k_t$ $R=0.6$ with $p_{\mathrm{T}}>20$\,GeV, numbers are jet $p_{\mathrm{T}}$ in GeV, ``H'' marks the $b$-jets from the Higgs boson. AlphaHits finds all six GenJets with the right flavour. Delphes finds five and tags three.}
  \label{fig:hf:event}
\end{figure}

\bibliographystyle{JHEP}
\bibliography{master_checked_references_v2}

\providecommand{\href}[2]{#2}\begingroup\raggedright\begin{thebibliography}{10}

\bibitem{barman2025largephysicsmodels}
K.G.~Barman, S.~Caron, E.~Sullivan, H.W.~de~Regt, R.~Ruiz~de Austri, M.~Boon
  et~al., \emph{Large Physics Models: Towards a collaborative approach with
  Large Language Models and Foundation Models},
  \href{https://doi.org/10.1140/epjc/s10052-025-14707-8}{\emph{Eur. Phys. J. C}
  {\bfseries 85} (2025) 1066}
  [\href{https://arxiv.org/abs/2501.05382}{{\ttfamily 2501.05382}}].

\bibitem{Caron2025Trackformers}
S.~Caron, N.~Dobreva, A.~Ferrer~S{\'a}nchez, J.D.~Mart{\'i}n-Guerrero,
  U.~Odyurt, R.~Ruiz~de Austri~Bazan et~al., \emph{Trackformers: in search of
  transformer-based particle tracking for the high-luminosity LHC era},
  \href{https://doi.org/10.1140/epjc/s10052-025-14156-3}{\emph{European
  Physical Journal C} {\bfseries 85} (2025) 460}
  [\href{https://arxiv.org/abs/2407.07179}{{\ttfamily 2407.07179}}].

\bibitem{Caron2025Trackformers2}
S.~Caron, N.~Dobreva, M.~Kimpel, U.~Odyurt, S.~Pshenov, R.~Ruiz~de Austri~Bazan
  et~al., \emph{TrackFormers Part 2: Enhanced Transformer-Based Models for
  High-Energy Physics Track Reconstruction},
  \href{https://arxiv.org/abs/2509.26411}{{\ttfamily 2509.26411}}.

\bibitem{Caron2025Higgsformer}
S.~Caron, P.~Moskvitina, R.~Ruiz~de Austri and E.~Shalugin, \emph{{Hits to
  Higgs: hit-level Higgs classification from raw LHC detector data using
  Higgsformer}},
  \href{https://doi.org/10.1140/epjc/s10052-026-15930-7}{\emph{Eur. Phys. J. C}
  {\bfseries 86} (2026) 848}
  [\href{https://arxiv.org/abs/2508.19190}{{\ttfamily 2508.19190}}].

\bibitem{Qu2022ParticleTransformer}
H.~Qu, C.~Li and S.~Qian, \emph{Particle Transformer for Jet Tagging},  in
  \emph{Proceedings of the 39th International Conference on Machine Learning},
  vol.~162 of \emph{Proceedings of Machine Learning Research},
  pp.~18281--18292, PMLR, 2022
  [\href{https://arxiv.org/abs/2202.03772}{{\ttfamily 2202.03772}}].

\bibitem{Kiehn2019TrackMLKaggle}
M.~Kiehn, S.~Amrouche, P.~Calafiura, V.~Estrade, S.~Farrell et~al., \emph{The
  TrackML high-energy physics tracking challenge on Kaggle},
  \href{https://doi.org/10.1051/epjconf/201921406037}{\emph{EPJ Web of
  Conferences} {\bfseries 214} (2019) 06037}.

\bibitem{golling2024mpm}
T.~Golling, L.~Heinrich, M.~Kagan, S.~Klein, M.~Leigh, M.~Osadchy et~al.,
  \emph{Masked particle modeling on sets: towards self-supervised high energy
  physics foundation models},
  \href{https://doi.org/10.1088/2632-2153/ad64a8}{\emph{Machine Learning:
  Science and Technology} {\bfseries 5} (2024) 035074}
  [\href{https://arxiv.org/abs/2401.13537}{{\ttfamily 2401.13537}}].

\bibitem{vigl2024finetuning}
M.~Vigl, N.~Hartman and L.~Heinrich, \emph{Finetuning foundation models for
  joint analysis optimization in High Energy Physics},
  \href{https://doi.org/10.1088/2632-2153/ad55a3}{\emph{Machine Learning:
  Science and Technology} {\bfseries 5} (2024) 025075}
  [\href{https://arxiv.org/abs/2401.13536}{{\ttfamily 2401.13536}}].

\bibitem{hsu2026evenet}
T.-H.~Hsu, B.-H.~Zhou, Q.~Liu, Y.~Xu, S.~Li, G.W.-S.~Hou et~al.,
  \emph{Foundation model for particle collision data analysis},
  \href{https://doi.org/10.1038/s42256-026-01322-9}{\emph{Nature Machine
  Intelligence} (2026) } [\href{https://arxiv.org/abs/2601.17126}{{\ttfamily
  2601.17126}}].

\bibitem{birk2024omnijet}
J.~Birk, A.~Hallin and G.~Kasieczka, \emph{{OmniJet-$\alpha$: The First
  Cross-Task Foundation Model for Particle Physics}},
  \href{https://doi.org/10.1088/2632-2153/ad66ad}{\emph{Machine Learning:
  Science and Technology} {\bfseries 5} (2024) 035031}
  [\href{https://arxiv.org/abs/2403.05618}{{\ttfamily 2403.05618}}].

\bibitem{mikuni2024omnilearn}
V.~Mikuni and B.~Nachman, \emph{{OmniLearn: A Method to Simultaneously
  Facilitate All Jet Physics Tasks}},
  \href{https://doi.org/10.1103/PhysRevD.111.054015}{\emph{Physical Review D}
  {\bfseries 111} (2025) 054015}
  [\href{https://arxiv.org/abs/2404.16091}{{\ttfamily 2404.16091}}].

\bibitem{bhimji2025omnilearned}
W.~Bhimji, C.~Harris, V.~Mikuni and B.~Nachman, \emph{{Foundation Model
  Framework for All Tasks Involving Jet Physics}},
  \href{https://doi.org/10.1103/knmd-f5jm}{\emph{Physical Review D} {\bfseries
  113} (2026) 032020} [\href{https://arxiv.org/abs/2510.24066}{{\ttfamily
  2510.24066}}].

\bibitem{amram2024aspen}
O.~Amram, L.~Anzalone, J.~Birk, D.A.~Faroughy, A.~Hallin, G.~Kasieczka et~al.,
  \emph{{Aspen Open Jets: Unlocking LHC Data for Foundation Models in Particle
  Physics}}, \href{https://doi.org/10.1088/2632-2153/ade58f}{\emph{Machine
  Learning: Science and Technology} (2025) }
  [\href{https://arxiv.org/abs/2412.10504}{{\ttfamily 2412.10504}}].

\bibitem{kakati2026prompting}
N.~Kakati, D.~Murnane, B.~Hashemi, S.~Klein, J.~Krupa, E.~Gross et~al.,
  \emph{Prompting Particle Physics: Tokenized Multi-modal Foundation Models for
  Combinatorially Many Tasks},  2026.
\newblock 10.48550/arXiv.2609.31862.

\bibitem{mokhtar2026mlpf}
F.~Mokhtar, J.~Pata, M.~Kagan and J.~Duarte, \emph{{Machine-Learned Particle
  Flow as a Foundation Model for Collider Physics}},
  \href{https://doi.org/10.1103/v5n3-d3tn}{\emph{PRX Intelligence} {\bfseries
  1} (2026) 023012} [\href{https://arxiv.org/abs/2606.14373}{{\ttfamily
  2606.14373}}].

\bibitem{hart2026rem}
M.~Hart, K.Y.~Wong, S.~Van~Stroud, P.~Duckett, T.~Scanlon and G.~Facini,
  \emph{{Reconstruct Everything Everywhere All at Once: An End-to-End Model for
  Future Colliders}},  \href{https://arxiv.org/abs/2610.41889}{{\ttfamily
  2610.41889}}.

\bibitem{Ju2021ExaTrkX}
X.~Ju, D.~Murnane, P.~Calafiura, N.~Choma, S.~Conlon, S.~Farrell et~al.,
  \emph{Performance of a geometric deep learning pipeline for HL-LHC particle
  tracking},
  \href{https://doi.org/10.1140/epjc/s10052-021-09675-8}{\emph{European
  Physical Journal C} {\bfseries 81} (2021) 876}
  [\href{https://arxiv.org/abs/2103.06995}{{\ttfamily 2103.06995}}].

\bibitem{Biscarat2021RealisticGNNTracking}
C.~Biscarat, S.~Caillou, C.~Rougier, J.~Stark and J.~Zahreddine, \emph{Towards
  a realistic track reconstruction algorithm based on graph neural networks for
  the HL-LHC}, \href{https://doi.org/10.1051/epjconf/202125103047}{\emph{EPJ
  Web of Conferences} {\bfseries 251} (2021) 03047}
  [\href{https://arxiv.org/abs/2103.00916}{{\ttfamily 2103.00916}}].

\bibitem{Caillou2024GNN4ITk}
S.~Caillou, P.~Calafiura, X.~Ju, D.~Murnane, T.~Pham, C.~Rougier et~al.,
  \emph{Physics Performance of the ATLAS GNN4ITk Track Reconstruction Chain},
  \href{https://doi.org/10.1051/epjconf/202429503030}{\emph{EPJ Web of
  Conferences} {\bfseries 295} (2024) 03030}.

\bibitem{Kieseler2020ObjectCondensation}
J.~Kieseler, \emph{Object condensation: one-stage grid-free multi-object
  reconstruction in physics detectors, graph, and image data},
  \href{https://doi.org/10.1140/epjc/s10052-020-08461-2}{\emph{European
  Physical Journal C} {\bfseries 80} (2020) 886}
  [\href{https://arxiv.org/abs/2002.03605}{{\ttfamily 2002.03605}}].

\bibitem{Lieret2023HighPileupOC}
K.~Lieret, G.~DeZoort, D.~Chatterjee, J.~Park, S.~Miao and P.~Li, \emph{High
  Pileup Particle Tracking with Object Condensation},
  \href{https://arxiv.org/abs/2312.03823}{{\ttfamily 2312.03823}}.

\bibitem{Calafiura2024EggNet}
P.~Calafiura, J.~Chan, L.~Delabrouille and B.~Wang, \emph{EggNet: An Evolving
  Graph-based Graph Attention Network for Particle Track Reconstruction},
  2024.

\bibitem{Chan2025EggNetPerformance}
J.~Chan, B.~Wang and P.~Calafiura, \emph{Physics and Computing Performance of
  the EggNet Tracking Pipeline},  2025.

\bibitem{Melkani2024TrackSorter}
Y.~Melkani and X.~Ju, \emph{TrackSorter: A Transformer-based sorting algorithm
  for track finding in High Energy Physics},
  \href{https://arxiv.org/abs/2407.21290}{{\ttfamily 2407.21290}}.

\bibitem{VanStroud2025TransformerTracking}
S.~Van~Stroud, P.~Duckett, M.~Hart, N.~Pond, S.~Rettie, G.~Facini et~al.,
  \emph{Transformers for Charged Particle Track Reconstruction in High Energy
  Physics}, \href{https://doi.org/10.1103/md46-yqgd}{\emph{Physical Review X}
  {\bfseries 15} (2025) 041046}
  [\href{https://arxiv.org/abs/2411.07149}{{\ttfamily 2411.07149}}].

\bibitem{miao2026heptv2}
S.~Miao, S.~Govil, J.P.~Rodgers, M.~Liu, J.~Duarte, S.-C.~Hsu et~al.,
  \emph{{HEPTv2: End-to-End Efficient Point Transformer for Charged Particle
  Reconstruction}},  \href{https://arxiv.org/abs/2606.20437}{{\ttfamily
  2606.20437}}.

\bibitem{Dao2023FlashAttention2ICLR}
T.~Dao, \emph{FlashAttention-2: Faster Attention with Better Parallelism and
  Work Partitioning},  2023.

\bibitem{He2024FlexAttention}
J.~Dong, B.~Feng, D.~Guessous, Y.~Liang and H.~He, \emph{Flex Attention: A
  Programming Model for Generating Optimized Attention Kernels},  2024.

\bibitem{Miao2024HEPT}
S.~Miao, Z.~Lu, M.~Liu, J.~Duarte and P.~Li, \emph{Locality-Sensitive
  Hashing-Based Efficient Point Transformer with Applications in High-Energy
  Physics},  \href{https://arxiv.org/abs/2402.12535}{{\ttfamily 2402.12535}}.

\bibitem{Locatello2020SlotAttention}
F.~Locatello, D.~Weissenborn, T.~Unterthiner, A.~Mahendran, G.~Heigold,
  J.~Uszkoreit et~al., \emph{Object-Centric Learning with Slot Attention},  in
  \emph{Advances in Neural Information Processing Systems}, vol.~33, 2020
  [\href{https://arxiv.org/abs/2006.15055}{{\ttfamily 2006.15055}}].

\bibitem{Cheng2022Mask2Former}
B.~Cheng, I.~Misra, A.G.~Schwing, A.~Kirillov and R.~Girdhar,
  \emph{Masked-attention Mask Transformer for Universal Image Segmentation},
  in \emph{Proceedings of the IEEE/CVF Conference on Computer Vision and
  Pattern Recognition}, 2022
  [\href{https://arxiv.org/abs/2112.01527}{{\ttfamily 2112.01527}}].

\bibitem{Builtjes2025PhysicsInteractions}
L.~Builtjes, S.~Caron, P.~Moskvitina, C.~Nellist, R.~Ruiz~de Austri,
  R.~Verheyen et~al., \emph{Attention to the strengths of physical
  interactions: Transformer and graph-based event classification for particle
  physics experiments},
  \href{https://doi.org/10.21468/SciPostPhys.19.1.028}{\emph{SciPost Physics}
  {\bfseries 19} (2025) 028}
  [\href{https://arxiv.org/abs/2211.05143}{{\ttfamily 2211.05143}}].

\bibitem{Amrouche2020TrackML}
S.~Amrouche et~al., \emph{The Tracking Machine Learning Challenge: Accuracy
  Phase},  in \emph{The NeurIPS '18 Competition}, The Springer Series on
  Challenges in Machine Learning, pp.~231--264, Springer (2020),
  \href{https://doi.org/10.1007/978-3-030-29135-8_9}{DOI}
  [\href{https://arxiv.org/abs/1904.06778}{{\ttfamily 1904.06778}}].

\bibitem{shazeer2017outrageously}
N.~Shazeer, A.~Mirhoseini, K.~Maziarz, A.~Davis, Q.V.~Le, G.E.~Hinton et~al.,
  \emph{Outrageously Large Neural Networks: The Sparsely-Gated
  Mixture-of-Experts Layer}, {\emph{arXiv preprint arXiv:1701.06538} (2017) }.

\bibitem{kuhn1955hungarian}
H.W.~Kuhn, \emph{The Hungarian Method for the Assignment Problem}, {\emph{Naval
  Research Logistics Quarterly} {\bfseries 2} (1955) 83}.

\bibitem{carion2020end}
N.~Carion, F.~Massa, G.~Synnaeve, N.~Usunier, A.~Kirillov and S.~Zagoruyko,
  \emph{End-to-End Object Detection with Transformers},  in \emph{European
  Conference on Computer Vision}, pp.~213--229, 2020.

\bibitem{sudre2017generalised}
C.H.~Sudre, W.~Li, T.~Vercauteren, S.~Ourselin and M.J.~Cardoso,
  \emph{Generalised Dice Overlap as a Deep Learning Loss Function for Highly
  Unbalanced Segmentations},  in \emph{Deep Learning in Medical Image Analysis
  and Multimodal Learning for Clinical Decision Support}, pp.~240--248,
  Springer, 2017.

\bibitem{Jegou2011PQ}
H.~J{\'e}gou, M.~Douze and C.~Schmid, \emph{Product Quantization for Nearest
  Neighbor Search}, \href{https://doi.org/10.1109/TPAMI.2010.57}{\emph{IEEE
  Transactions on Pattern Analysis and Machine Intelligence} {\bfseries 33}
  (2011) 117}.

\bibitem{Mentzer2024FSQ}
F.~Mentzer, D.~Minnen, E.~Agustsson and M.~Tschannen, \emph{Finite Scalar
  Quantization: {VQ-VAE} Made Simple},  in \emph{International Conference on
  Learning Representations}, 2024
  [\href{https://arxiv.org/abs/2309.15505}{{\ttfamily 2309.15505}}].

\bibitem{vandenOord2017VQVAE}
A.~van~den Oord, O.~Vinyals and K.~Kavukcuoglu, \emph{Neural Discrete
  Representation Learning},  in \emph{Advances in Neural Information Processing
  Systems}, vol.~30, 2017 [\href{https://arxiv.org/abs/1711.00937}{{\ttfamily
  1711.00937}}].

\bibitem{Mitzenmacher2017Probability}
M.~Mitzenmacher and E.~Upfal, \emph{Probability and Computing: Randomization
  and Probabilistic Techniques in Algorithms and Data Analysis}, Cambridge
  University Press, 2nd~ed. (2017).

\bibitem{Lancucki2020RobustVQ}
A.~{\L}a{\'n}cucki, J.~Chorowski, G.~Sanchez, R.~Marxer, N.~Chen,
  H.J.G.A.~Dolfing et~al., \emph{Robust Training of Vector Quantized Bottleneck
  Models},  in \emph{International Joint Conference on Neural Networks
  (IJCNN)}, 2020, \href{https://doi.org/10.1109/IJCNN48605.2020.9207145}{DOI}
  [\href{https://arxiv.org/abs/2005.08520}{{\ttfamily 2005.08520}}].

\bibitem{Zhu2024VQGANLC}
L.~Zhu, F.~Wei, Y.~Lu and D.~Chen, \emph{Scaling the Codebook Size of {VQGAN}
  to 100,000 with a Utilization Rate of 99\%},  in \emph{Advances in Neural
  Information Processing Systems}, vol.~37, 2024
  [\href{https://arxiv.org/abs/2406.11837}{{\ttfamily 2406.11837}}].

\bibitem{Delphes2014}
{\scshape DELPHES 3} collaboration, \emph{{DELPHES} 3, A modular framework for
  fast simulation of a generic collider experiment},
  \href{https://doi.org/10.1007/JHEP02(2014)057}{\emph{Journal of High Energy
  Physics} {\bfseries 2014} (2014) 57}
  [\href{https://arxiv.org/abs/1307.6346}{{\ttfamily 1307.6346}}].

\end{thebibliography}\endgroup

\end{document}